\documentclass[twocolumn,prd,nofootinbib,superscriptaddress,eqsecnum,tightenlines,8pt]{revtex4} 

\usepackage{fancyhdr}
\usepackage{amsmath,amsfonts,amssymb}

\usepackage{color}
\definecolor{red}{rgb}{1,0,0}
\definecolor{gre}{rgb}{0,0.6,0}
\definecolor{blu}{rgb}{0,0,1}

\usepackage[pdftex]{graphicx}
\usepackage{mathrsfs}
\DeclareGraphicsExtensions{.jpg, .JPG , .png , .pdf, .gif}

\newcommand{\lb}[1]{\label{#1}}
\newcommand{\ff}[1]{\ref{#1}}

\newcommand{\bark}{\bar{k}}
\newcommand{\barp}{\bar{p}}
\newcommand{\barmu}{\bar{\mu}}
\newcommand{\barn}{\bar{N}}

\newcommand{\barN}{\bar{N}}

\newcommand{\barnu}{\bar{\nu}}
\newcommand{\barpi}{\bar{\pi}}

\newcommand{\K}[1]{\mathbb{K} \left[#1 \right]}

\newcommand{\Hc}{\mathscr{H}}

\begin{document}

\title{Anomaly-free perturbations with inverse-volume and holonomy corrections \\ 
in loop quantum cosmology}

\date{\today}

\author{Thomas Cailleteau}
 \email{thomas@gravity.psu.edu}
\affiliation{%
Institute for Gravitation \& the Cosmos, Penn State\\
University Park, PA 16802, USA
}

\author{Linda Linsefors}%
 \email{linsefors@lpsc.in2p3.fr}
\affiliation{%
Laboratoire de Physique Subatomique et de Cosmologie, UJF, INPG, CNRS, IN2P3\\
53, avenue des Martyrs, 38026 Grenoble cedex, France
}%

\author{Aurelien Barrau}%
 \email{Aurelien.Barrau@cern.ch}
\affiliation{%
Laboratoire de Physique Subatomique et de Cosmologie, UJF, INPG, CNRS, IN2P3\\
53,avenue des Martyrs, 38026 Grenoble cedex, France\\
and\\
Institut des Hautes Etudes Scientifiques,\\
35 route de Chartres, 91440 Bures-sur-Yvette, France
}%

\begin{abstract}
This article addresses the issue of the closure of the algebra of constraints for generic (cosmological) perturbations when taking into account 
simultaneously the two main corrections of effective loop quantum cosmology, namely the holonomy and the 
inverse-volume terms. Previous works on either the holonomy or the inverse-volume case are reviewed and 
generalized. In the inverse-volume case, we point out new possibilities. An anomaly-free solution including both corrections is found for perturbations, and the corresponding equations of motion are derived.
\end{abstract}

\maketitle

%\tableofcontents

\section{Introduction}
Loop quantum gravity (LQG) is a tentative theory of quantum geometry that builds on both Einstein gravity and quantum physics
without any fundamental new principle. Reviews can be found in \cite{lqg_review}. Loop Quantum Cosmology (LQC) is basically the
symmetry reduced version of LQG (see \cite{lqc_review} for general introductions to LQC). At this stage a rigorous derivation of LQC from the
full mother theory is not yet possible. Rather, LQC mostly imports the main techniques of LQG in the cosmological sector. It relies on 
a kinematical Hilbert space that is different than in the Wheeler-DeWitt case. This so-called polymeric quantization has been
shown to be unique when diffeomorphism invariance is rigorously imposed \cite{unique}. In this space, in order for the quantum version of the constraints to be well defined, an operator is
associated not with the connection, but only with its holonomy. The basic variables of LQC are therefore the holonomy of the
connection and its conjugate variable, the flux of the densitized triads.\\

%At the effective level, LQC can be modeled by two kind of corrections. 
At the effective level, it is believed that LQC can
be modeled by two main corrections. In this framework, we restrain our
study to an isotropic and homogeneous universe, 
considering neither possible backreaction effects  \cite{backreaction}, nor higher
derivatives terms in the constraints via momentum terms \cite{momentumterms} . Consequently,
the quantum corrections studied here  are the simplest homogeneous and isotropic corrections one can use.  
The inverse-volume correction \cite{iv} (or inverse-triad
if one
relaxes the isotropy hypothesis) provides natural cut-off functions of divergences for factors containing inverse components of
densitized triads, arising from spatial discreteness. The holonomy correction \cite{holo} is instead associated with higher 
powers of 
intrinsic and extrinsic spatial curvature components stemming from the appearance of holonomies of the Ashtekar connection.\\

%Here, we focus on the well known problem of the consistency of the effective theory \red{in an homogeneous and isotropic universe}. This means that the evolution
%produced by the model should be consistent with the theory itself.
Here, we focus on the well known problem of the consistency of the effective theory of perturbations around an homogeneous and isotropic universe. By consistency, we mean that the evolution
produced by the model should be consistent with the theory itself. This basically translates into the requirement that the
Poisson bracket between two constraints should be proportional to another constraint. The coefficient of proportionality being a
function of the fundamental variables, this makes the situation slightly more subtle than in usual field theories dealing with
simple structure constants. Another important point is that the closure of the algebra of constraints \cite{Dirac} should even be considered off-shell \cite{bojo1}.\\

To some extent, this article is more technical than physical. The physical consequences will need more work to be fully explored.  The aim of this work is not to describe all the steps of the derivations, but rather to give hints on the way it has been done. More information can also be found in the references given.
Here, we first re-address the case of inverse-volume corrections. Previous results are revised. Then a reminder on
the holonomy-corrected algebra is presented. Finally, the inverse-volume and holonomy case is studied and solved.

\section{LQC perturbations}

In the canonical formulation of general relativity, the Hamiltonian
is a sum of three constraints: 
\begin{equation}
H[N^i,N^a,N] = \frac{1}{2\kappa } \int_{\Sigma}d^3x \left( N^i C_i+N^a C_a+N C\right) \approx 0,
\nonumber 
\end{equation}
where $\kappa = 8\pi G$, $(N^i,N^a,N)$ are Lagrange multipliers,  $C_i$ is 
the Gauss constraint,  $C_a$ is the diffeomorphism  constraint, and $C$ is 
the scalar constraint.  The sign "$\approx$" means in this context equality on the surface of 
constraints. The smeared 
constraints, whose expressions  will be given later, are defined so that $H[N^i,N^a,N] = G[N^i]+D[N^a]+H[N]$: 
\begin{eqnarray}
\mathcal{C}_1 &=& G[N^i] =\frac{1}{2\kappa } \int_{\Sigma}d^3x\ N^i C_i, \\
\mathcal{C}_2 &=& D[N^a] =\frac{1}{2\kappa } \int_{\Sigma}d^3x\ N^a C_a, \\
\mathcal{C}_3 &=& H[N] =\frac{1}{2\kappa } \int_{\Sigma}d^3x\ N C.
\end{eqnarray}
The full Hamiltonian is a total constraint which is vanishing for all multiplier functions $(N^i,N^a,N)$. 

As $H[N^i,N^a,N] \approx 0$ at any times, the time derivative
of the Hamiltonian constraint is also weakly vanishing, $\dot{H}[N^i,N^a,N] \approx 0$. 
Due to the Hamilton equation $\dot{f}=\{f,H[M^i,M^a,M]\}$, one has
\begin{equation}
\left\{H[N^i,N^a,N], H[M^i,M^a,M]\right\} \approx 0. \label{HH}
\end{equation}
Because of the linearity of the Poisson brackets, it is straightforward to find that the condition 
(\ref{HH}) is fulfilled if the smeared constraints belong to a first class algebra
\begin{equation}
\{ \mathcal{C}_I, \mathcal{C}_J \} = {f^K}_{IJ}(A^j_b,E^a_i) \mathcal{C}_K, \label{algebra}
\end{equation}
also called "algebra of deformation".
In Eq. (\ref{algebra}), the ${f^K}_{IJ}( A^j_b,E^a_i)$ are structure functions which, in general, 
depend on the phase space (Ashtekar) variables  $(A^j_b, E^a_i)$. The algebra 
of constraints is fulfilled at the classical level due to general covariance. 
To prevent the system from escaping the surface of constraints, leading to an
unphysical behavior, the algebra must also be closed at the quantum level.  
In addition, as stated in the introduction and  as pointed out in \cite{Nicolai:2005mc}, 
the algebra of quantum constraints should be strongly closed (that is should
close \emph{off shell}). 
This means that the relation (\ref{algebra}) should hold in
the whole kinematical phase space, and not only on the surface of constraints. This should 
remain true after promoting the 
constraints to quantum operators.  

In effective LQC, the constraints are quantum-modified and the corresponding 
Poisson algebra might not be closed:  
\begin{equation}
\{ \mathcal{C}^Q_I, \mathcal{C}^Q_J \} = {f^K}_{IJ}(A^j_b,E^a_i) \mathcal{C}^Q_K+
\mathcal{A}_{IJ},
\end{equation}
where $\mathcal{A}_{IJ}$ stands for the anomaly term which can appear due to the 
quantum modifications. For consistency (closure of algebra), $\mathcal{A}_{IJ}$ is 
required to vanish. The condition $\mathcal{A}_{IJ}=0$ induces some restrictions on 
the form of the quantum corrections. More importantly this requires the addition of counterterms
that should vanish at the classical limit.
 
The question of the construction of an anomaly-free algebra of constraints is especially interesting to 
address in inhomogeneous LQC. Perturbations around 
the cosmological background are indeed responsible for structure formation in the Universe and
are a promising way to try to test the theory. In the case of a flat FLRW background, the Ashtekar
variables can be decomposed as 
\begin{equation}
 A^i_a = \gamma \bar{k} \delta^i_a +\delta A^i_a \ \ \ {\rm and} \ \ \ E_i^a = \bar{p} \delta_i^a +\delta E_i^a, 
\end{equation}
where $\bar{k}$ and $\bar{p}$ parametrize the background phase space, and 
$\gamma$ is the so-called Barbero-Immirzi parameter. The issue of anomaly freedom for the algebra of cosmological perturbations was extensively 
studied for  inverse-volume corrections. It was shown that this requirement
can be fulfilled for first order perturbations. 
This was derived for scalar \cite{bojo1,Bojowald:2008jv}, 
vector \cite{Bojowald:2007hv} and tensor perturbations \cite{Bojowald:2007cd}. 
Based on the anomaly-free scalar perturbations, predictions 
for the power spectrum of cosmological perturbations were also performed 
\cite{Bojowald:2010me}. This gave a chance to put constraints on 
some parameters of the model using observations of the cosmic microwave 
background radiation (CMB) \cite{Bojowald:2011hd}. The closure for holonomy-corrected vector
modes was obtained in \cite{vectors} and for holonomy-corrected scalar modes
in \cite{scalars} and \cite{WilsonEwing:2012bx}. Interestingly, it was pointed out that the requirement of consistency for scalar
modes also modifies the algebra of tensor modes \cite{consistency}, leading to a possible effective
change from hyperbolic to elliptic regime around the bounce with this deformed algebra \cite{deformed}. It was also  found recently that under a change of variables, the usual non-deformed algebra of general relativity could be recovered \cite{Tibrewala:2013kba}.\\

The aim of this article is to address the issue of anomaly freedom when both holonomy and inverse-volume
corrections are included. Our path is not the only possible one. In \cite{abhay1}, several 
criticisms were formulated. However it seems to us that, at this stage, this deformed
algebra approach (see \cite{deformed} for a review) is suitable enough to capture the main features
of the full theory at an effective level \cite{Rovelli:2013zaa}.

\section{Inverse-volume case}

Whenever inverse powers of densitized triads components, that would classically diverge at
singularities, appear, an effective quantum correction, called "inverse-volume", is expected. A
comprehensive treatment was performed in \cite{bojo1}. However, to prepare for the simultaneous
treatment of both holonomy and inverse-volume terms, we re-visit and generalize this issue. 
In \cite{bojo1}, where only inverse-volume corrections were studied, the authors basically considered only linear terms in the anomalies. However, in the case of holonomy corrections, it has been shown \cite{scalars} that a more rigorous treatment can be done by solving the exact non-linear anomalies. Therefore, in the following, even for the inverse-volume term alone, the anomalies will not be approximated and their non-linear properties will be used to derive the general expressions for the counterterms. This is consistent with the holonomy case. 
  \\

In the \textit{semi-classical limit}, where quantum corrections are small, the inverse-volume correction can be expressed as \cite{bojowald_inverse_volume}
\begin{equation} \lb{evolIV}
\gamma_0 (\barp) =  1 + \frac{\lambda }{\barp^{n}},
\end{equation}
where $n$ is assumed to be positive. This is a meaningful assumption as long as one considers scales of inhomogeneities that are larger than the scale of discreetness. In the following, we will however try to find the expression of the counterterms whatever the behavior of the inverse-volume correction. We will therefore assume no particular expression and see that in order to close the algebra, this correction will have to fulfill some differential equations in agreement with what was obtained at the semi-classical limit.

In order to extract and resolve the anomalies, we will first express the modified constraints and add counterterms $\alpha_i$ and $\beta_i$ in their expressions. We are here interested in the linear theory of perturbations, so the constraints will be considered up to the second order in perturbations only.
Moreover, it has been shown previously \cite{consistency} that the case of scalar perturbations is the more general one. For this reason we will focus on this one. \\
 
In LQG, the diffeomorphism constraints are expected not to be modified by the quantum corrections. In the gravitational
sector, this leads to:
\begin{equation}\lb{DG}
D_G[N^c]\!\! = \!\! \int \frac{d^3x}{\kappa} \delta N^c \left[ \barp \partial_c \delta K^d_d -
\barp \partial_d \delta K^d_c - \bark \partial_d \delta E^d_c \right],
\end{equation}
where spatial and internal indices are set identically for clarity only, \textit{i.e.} $\delta K^d_d \dot{=} \delta K^i_a \delta^a_i$ as they won't play any role in this study.
The Hamiltonian constraint reads as
\begin{eqnarray}\lb{conttot}
H^Q_G[N] &=&  \frac{1}{2 \kappa} \int d^3x (\barN + \delta N) (\gamma_0+  \gamma^{(1)}+ \gamma^{(2)}) \nonumber \\
&& \times \left[ \mathcal{H}^{(0)}_G +\mathcal{H}^{(1)}_G + \mathcal{H}^{(2)}_G \right].
\end{eqnarray}
Here, $\gamma_0$ is the homogeneous and isotropic part of the inverse-volume correction whereas $\gamma^{(1)}$ and $\gamma^{(2)}$ are the inhomogeneous parts, at first and second order in $\delta E$ and $\delta K$. In the classical limit: $\gamma_0=1$, $\gamma^{(1)}=\gamma^{(2)}=0$.

The zeroth, first and second orders of the gravitational Hamiltonian densities
are:

\begin{eqnarray}
&&\mathcal{H}^{(0)}_G = - 6 \sqrt{\barp} \bark^2, \\
&&\mathcal{H}^{(1)}_G = - 4 \sqrt{\barp}( \bark + \alpha_1 )\delta K^d_d  -  \frac{1}{\sqrt{\barp}}  \left(\bark^2+ \alpha_2 \right) \delta E^d_d \nonumber \\
&& +  \frac{2}{\sqrt{\barp}}\left(1 + \alpha_3 \right) \partial_c \partial^d \delta E^c_d,  \\
&&\mathcal{H}^{(2)}_G=\sqrt{\barp} \left(1 + \alpha_4\right) \delta K^d_c \delta K^c_d- \sqrt{\barp} \left(1 + \alpha_5\right) (\delta K^d_d)^2   \nonumber \\ 
&&-   \frac{2}{\sqrt{\barp}} \left(\bark + \alpha_6 \right) \delta E^c_d \delta K^d_c  - \frac{1}{2 \barp^{\frac{3}{2}}}  \left( \bark^2+ \alpha_7  \right) \delta E^c_d \delta E^d_c  \\
&& + \frac{1}{4 \barp^{\frac{3}{2}}} \left(\bark^2 +\alpha_8 \right) (\delta E^d_d)^2 - \frac{\delta^{jk}}{2 \barp^{\frac{3}{2}}} \left(1 + \alpha_9 \right) (\partial_c \delta E^c_j) (\partial_d \delta E^d_k).\nonumber
\end{eqnarray}

The very same approach should be applied to the matter sector. We assume through out the paper that the matter consists of a single scalar field with a potential $V(\phi)$. It is interesting to notice that the approach used here leads to the same counterterms, whatever the chosen shape of the potential.

The diffeomorphism constraint is
\begin{eqnarray}\lb{Dm}
D_m[N^c]\!\! = \!\! \int d^3x \delta N^c \barpi \partial_c \delta \varphi,
\end{eqnarray}
and the Hamiltonian is
\begin{eqnarray}\lb{Hm}
&& H_m[N] = \int d^3x \barN \left[\nu_0 \mathcal{H}^{(0)}_{\pi} +\mathcal{H}^{(0)}_{\varphi}  \right] \\
&& + \int d^3x \delta N \left[ \nu^{(1)} \mathcal{H}^{(0)}_{\pi} + \nu_0 \mathcal{H}^{(1)}_{\pi} + \mathcal{H}^{(1)}_{\varphi} \right] \nonumber \\
&& + \int d^3x \barN \left[ \nu^{(2)} \mathcal{H}^{(0)}_{\pi} + \nu^{(1)} \mathcal{H}^{(1)}_{\pi} + \nu_0 \mathcal{H}^{(2)}_{\pi} + \sigma_0 \mathcal{H}^{(2)}_{\nabla} + \mathcal{H}^{(2)}_{\varphi}  \right]. \nonumber
\end{eqnarray}
The $\nu$ and $\sigma$ terms play  roles similar to the ones of the $\gamma$ terms in the gravitational sector. The different parts of the matter Hamiltonian read as:

\begin{eqnarray}\lb{constmatter}
\mathcal{H}^{(0)}_{\pi} &=& \frac{\barpi^2}{2 \barp^{\frac{3}{2}}}, \hskip0.5truecm \mathcal{H}^{(0)}_{\nabla} = 0, \hskip 0.5truecm \mathcal{H}^{(0)}_{\varphi} = \barp^{\frac{3}{2}} V, \\
\mathcal{H}^{(1)}_{\pi} &=& \frac{\barpi}{\barp^{\frac{3}{2}}}\left( 1 + \beta_1 \right) \delta \pi -  \frac{\barpi^2}{2 \barp^{\frac{3}{2}}} \left(1+ \beta_2 \right) \frac{\delta E^d_d }{2 \barp}, \lb{beta1const} \\
\mathcal{H}^{(1)}_{\nabla} &=& 0, \\
\mathcal{H}^{(1)}_{\varphi} &=&   \barp^{\frac{3}{2}} \partial_{\varphi} V\left(1 + \beta_3 \right) \delta \varphi +  \barp^{\frac{3}{2}} V \left( 1+ \beta_4 \right)\frac{\delta E^d_d}{2 \barp}, \lb{beta3const}\\
\mathcal{H}^{(2)}_{\pi} &=&   \frac{1}{ 2 \barp^{\frac{3}{2}}}\left(1 + \beta_5 \right) (\delta \pi)^2- \frac{\barpi}{ \barp^{\frac{3}{2}}}\left( 1 + \beta_6 \right)  \delta \pi \frac{\delta E^d_d}{2 \barp}  \\
&& + \frac{\barpi^2}{ 2 \barp^{\frac{3}{2}}} \left(1+ \beta_7  \right) \frac{(\delta E^d_d)^2}{8 \barp^2} + \frac{\barpi^2}{ 2 \barp^{\frac{3}{2}}} \left(1+ \beta_8  \right) \frac{\delta E^c_d \delta E^d_c }{4 \barp^2}, \nonumber\\
\mathcal{H}^{(2)}_{\nabla} &=&  \frac{\sqrt{\barp}}{2}\left(1  + \beta_9 \right) \partial_a \delta \varphi \partial^a \delta \varphi, \\
\mathcal{H}^{(2)}_{\varphi} &=& \frac{\barp^{\frac{3}{2}}}{2} \partial_{\varphi \varphi} V\left(1 + \beta_{10} \right) (\delta \varphi)^2 + \barp^{\frac{3}{2}} \partial_\varphi V \left( 1+ \beta_{11} \right) \delta \varphi \frac{\delta E^d_d}{2 \barp} \nonumber \\
&& +  \barp^{\frac{3}{2}} V \left(1+ \beta_{12} \right) \frac{(\delta E^d_d)^2}{8 \barp^2}  -\barp^{\frac{3}{2}} V  \left( 1+ \beta_{13} \right)  \frac{\delta E^c_d \delta E^d_c }{4 \barp^2}. \nonumber \\ \lb{constmatterEnd}
\end{eqnarray}

The counterterms $\alpha_i$ and $\beta_i$ are function of the homogeneous background.
In the spirit of LQG where matter fields live on top of the gravitational field, we assume that both the gravitational counterterms and the matter ones do not depend on the matter fields themselves but only on $\bark$ and $\barp$. This assumption can be questioned but is reasonable at this stage of  development of LQC. In the following, we will  therefore consider
\begin{eqnarray}
\alpha_i=\alpha_i[\barp,\bark],\\
\beta_i=\beta_i[\barp,\bark],
\end{eqnarray}
where $i \in N$ labels the counterterms and is not related to the indices of the phase space variables.

In this work, we have considered that it should be also the case for the inverse-volume correction (this will also remain true for the holonomy correction), whose dependence should only be in terms of the gravitational variables, namely $\bark$ and  $\barp$. It is therefore decomposed according to:

\begin{eqnarray}
\gamma_0 &=& \gamma_0[\barp,\bark], \\
\gamma^{(1)} &=& \gamma_1 \delta E^d_d + \gamma_2 \delta K^d_d, \\
\gamma^{(2)} &=& \gamma_3 \delta E^c_d \delta E^d_c + \gamma_4 (\delta E^d_d)^2 \nonumber \\
&& +\gamma_5 \delta K^c_d \delta K^d_c  + \gamma_6 (\delta K^d_d)^2 \nonumber \\
&& +\gamma_7 \delta E^c_d \delta K^d_c + \gamma_8 \delta E^d_d \delta K^c_c ,
\end{eqnarray}
and
\begin{eqnarray}
\nu_0 &=& \nu_0[\barp,\bark], \\
\nu^{(1)} &=& \nu_1 \delta E^d_d + \nu_2 \delta K^d_d, \\
\nu^{(2)} &=& \nu_3 \delta E^c_d \delta E^d_c + \nu_4 (\delta E^d_d)^2 \nonumber \\
&& +\nu_5 \delta K^c_d \delta K^d_c  + \nu_6 (\delta K^d_d)^2\nonumber \\
&& +\nu_7 \delta E^c_d \delta K^d_c + \nu_8 \delta E^d_d \delta K^c_c , 
\end{eqnarray}
where all $\gamma_i=\gamma_i[\barp,\bark]$ and $\nu_i=\nu_i[\barp,\bark]$.

This parametrization can however be much simplified: it can be shown easily that, after replacing the former expressions in the constraint densities, $\gamma_3$, $\gamma_4$, $\gamma_5$, $\gamma_6$, $\gamma_7$, $\nu_1$, $\nu_3$, and $\nu_4$ can be absorbed into the counterterms. This setting will have no physical consequences, only technical ones.
In the case where the holonomy correction was considered with scalar perturbations \cite{scalars}, some unknown parameters were introduced in the definition of the correction, but after some calculations it was demonstrated that the correct counterterms will anyway compensate for these parameters, leading to physical results not  depending on them. The very same thing happens here. 
We will therefore choose $\gamma_3=\gamma_4=\gamma_5=\gamma_6=\gamma_7=\nu_1=\nu_3=\nu_4=0$ without loss of generality.

We should also notice that $\sigma_0$  appears only once together with $(1+\beta_9)$ in Eq. (\ref{Hm}). This means that only the combination $\sigma_0(1+\beta_9)$ is of physical importance and not the individual factors. Because of this, we can absorb the fraction $\frac{\sigma_0}{\nu_0}$ into $(1+\beta_9)$, that is we can choose $\sigma_0=\nu_0$ in Eq. (\ref{Hm}).
\\

\subsection{Classical limit}
When trying to close the algebra, it is mandatory to ensure that the classical limit is correct. This means that, when taking the limit \begin{equation}
\barp\to\infty, 
\end{equation}
the final expressions for the constraints should reduce to the classical ones.
Inverse-volume corrections are indeed assumed to be important only for small volumes, \textit{i.e.} for small $\barp=\bar{a}^2$, where $\bar{a}$ is the scale factor of the homogeneous part of the metric.

The interpretation of this limit is not without  problems. One can simply rescale $\bar{a}$, and therefore $\barp$, by a change of coordinates which is pure gauge. But such a gauge transformation will also affect the other quantities. Roughly speaking, if one decides to apply this change of coordinate, some of the quantities among the observables will remain unchanged, scale invariant, when taking the limit $\barp\to\infty$, but some others will be consequently diluted. In fact, the universe can tend to a Minkowski space just by such a transformation. 
%For the homogenius background:
%The observables are: H, \varphi, \dot{\varphi}
%Tha canonical cordinates are: \bark, \barp, \varphi, \pi
%
%The observables must remain unchanged under a cordinate transformation. \barp will ofcouse change if we rescale the coordinates. And from the fact that H and \dot{\varphi} remain unchanged, we can (at least in principle) derive how \bark and \pi change under such a transformation.
%
%However, we can choose to take the limit \barp->infty, while keeping \bark, \varphi, and \pi fixed. This this transformation will then NOT be pure gauge, and therfore the physical quantities will be affected.
%More exactly H->0, \dot{\varphi}->0 and \varphi remain unchanged.
%
%However the above transformation does not keep the solution on the constraint surface. But since the corrections and counter therms does not depend on the matter part, we do not actually have to care about their behaviour. So we are really taking the limit, \barp->infty while keeping only \bark fixed and staying on the constraint sruface. Than we get H->0 and \rho->0.
We do not expect quantum effects in Minkowski space.We therefore assume that, at the classical limit, $\forall (i,n) \in N^2$
\begin{eqnarray}
\gamma_0,\nu_0,\sigma_0&\to&1,\\
\gamma^{(1)},
\gamma^{(2)},
\nu^{(1)},
\nu^{(2)}&\to&0, \\
\text{$ \partial^n \gamma_i, \partial^n \nu_i, \partial^n \sigma_i \to 0, $}
\end{eqnarray}
and demand that
\begin{eqnarray}
\alpha_i&\to&0, \\
\beta_i&\to&0, \\
\text{$ \partial^n \alpha_i, \partial^n \beta_i\to 0, $}
\end{eqnarray}
when  $\barp\to\infty$.

\subsection{Calculation of the Poisson brackets}

This section is devoted to the calculation of the different Poisson brackets involved in the final algebra. In order to maintain some clarity in the article, we define:
\begin{eqnarray} \lb{lambdagamma}
\Lambda_{\gamma} &=& \frac{\barp}{\gamma_0} \frac{\partial \gamma_0}{\partial \barp}, \\ 
\Lambda_{\nu} &=& \frac{\barp}{\nu_0} \frac{\partial \nu_0}{\partial \barp}.\lb{lambdanu}
\end{eqnarray}

\subsubsection{$\{ H_G[N], D_G[N^c] \} $}

After some calculations, the Poisson bracket can be decomposed as a sum of independent terms.

\begin{eqnarray}
 && \{ H_G[N], D_G[N^c] \} =  - H_G[\delta N^c \partial_c \delta N] \lb{HGDGH}\\
 && + \gamma_0\frac{\barn}{\sqrt{\barp}}  \left(\bark \alpha_4 +\alpha_6 -\frac{\bark^2}{\gamma_0} \frac{\partial \gamma_0}{\partial \bark} \right) D_G[N^c]\lb{HGDGD} \\
 && - \int \frac{d^3x}{ \kappa}  \sqrt{\barp} \gamma_0 (\delta N \partial_c \delta N^c) \mathcal{A}_1 \lb{HGDG1}\\
 && + \int \frac{d^3x}{\kappa} \bark  \sqrt{\barp} \gamma_0  \barN (\delta N^c \partial_c \delta K^d_d) \mathcal{A}_2 \lb{HGDG2}\\
 && -  \int \frac{d^3x}{2 \kappa} \frac{\barn}{\sqrt{\barp}} \gamma_0 (\partial_c \delta N^c) (\delta E^d_d) \mathcal{A}_3 \lb{HGDG3} \\
&&+  \int \frac{d^3x}{2 \kappa} \frac{\barn}{\sqrt{\barp}} \gamma_0 (\partial_d \delta N^c) (\delta E^d_c) \mathcal{A}_4 \lb{HGDG4} \\
 && + \int \frac{d^3x}{\kappa} \sqrt{\barp}\barN (\partial_d \delta N^d ) (\partial^a \partial_c \delta E^c_a)\mathcal{A}_{5}  \lb{HGDG41}\\
 && + 2 \int \frac{d^3x}{\kappa} \sqrt{\barp} \gamma_0  (\partial_c \partial^d \delta N)(\partial_d \delta N^c - (\partial_a \delta N^a )\delta ^c_d )  \left(1 + \alpha_3 \right) \nonumber\\
 && \lb{HGDG5}\\
 && + \int \frac{d^3x}{\kappa} \sqrt{\barp} \gamma_0  (\partial_d \delta N^c - (\partial_a \delta N^a )\delta ^c_d )(\partial^d \partial_a \delta E^a_c) \left(1+ \alpha_9 \right). \nonumber\\
&& \lb{HGDG6}
\end{eqnarray}

From this expression it can be noticed that:

\begin{itemize}
\item Eq. (\ff{HGDGH}) is simply the classical, anomaly-free, expression.
\item Eq. (\ff{HGDGD}) is proportional to the gravitational diffeomorphism constraint which is an element of the algebra. However, it is not expected at the classical level and could be interpreted as an anomaly. Nevertheless, only the full algebra (taking also into account matter) is physically relevant. If the matter Poisson brackets exhibit a similar term  with the matter diffeomorphism constraint, it might then not be an anomaly. At this stage, this term should be kept.
\item $\mathcal{A}_{1}$, $\mathcal{A}_{2}$, $\mathcal{A}_{3}$, $\mathcal{A}_{4}$ and $\mathcal{A}_{5}$ are anomalies that must vanish. In other words, to close this part of the algebra we must have
\begin{equation}
\mathcal{A}_{1}=\mathcal{A}_{2}=\mathcal{A}_{3}=\mathcal{A}_{4}=\mathcal{A}_{5}=0.
\end{equation}
\item By integrating by part and due to the commutation property of the derivatives, it can be easily shown that Eqs. (\ff{HGDG5}) and (\ff{HGDG6}) vanish for any values of $\alpha_3$ and $\alpha_9$. This is true here for the scalar perturbations, but it remains correct  for any kind of perturbations \cite{consistency}.

\item Moreover, 
\begin{equation}
\mathcal{A}_{5}=\left(1 + \alpha_3 \right) \left(2\gamma_1+\frac{\bark}{\barp}\gamma_2\right)
\end{equation}
exhibits the term  $(1+\alpha_3)$.  However, we cannot demand that $(1+\alpha_3)=0$ as this would violate the classical limit. Therefore, we must have
\begin{equation}\lb{g1}
\gamma_1 = -\frac{\bark}{2\barp}\gamma_2,
\end{equation}
in order to cancel the anomaly. 

\end{itemize}

After using Eq. (\ref{g1}), the remaining anomalies lead to:
\begin{eqnarray}
\lb{A1}
\mathcal{A}_1 &=& 2 \bark \alpha_1 + \alpha_2,\\ \nonumber \\
\lb{A2}
\mathcal{A}_2 &=& \alpha_5- \alpha_4 \nonumber\\
&&+\frac{1}{\gamma_0}\left[\left(3\bark+2\alpha_1+\frac{\alpha_2}{\bark}\right)\gamma_2+6\bark\barp\gamma_8\right],\\ \nonumber \\
 \lb{A3}
\mathcal{A}_3 &=& \alpha_7-\alpha_8 \nonumber\\
&&+\frac{1}{\gamma_0}\left[-\left(3\bark^3+2\bark^2\alpha_1+\bark\alpha_2\right)\gamma_2+6\bark^3\barp\gamma_8\right],\\ \nonumber \\
\lb{A4}
\mathcal{A}_4 &=& -2\bark^2\alpha_4-4\bark\alpha_6+\alpha_7+2\bark^2\Lambda_\gamma+\frac{2\bark^3}{\gamma_0}\frac{\partial\gamma_0}{\partial{\bark}}. \qquad.
\end{eqnarray}

\subsubsection{$\{ H_G[N_1], H_G[N_2] \} $}

The computation of this Poisson bracket leads to:

\begin{eqnarray}
&& \{ H_G[N_1], H_G[N_2] \} \nonumber \\
&& = \gamma_0^2 (1+\alpha_3)(1+\alpha_4)  D_G\left[ \frac{\barN}{\barp} \partial^c (\delta N_2 - \delta N_1) \right]\lb{HGHGD}  \,\,\, \,\,\,\\ 
&& + \int \frac{d^3x}{\kappa} \gamma_0^2  \barn \Delta (\delta N_2 -\delta N_1)(\delta K^d_d)  \mathcal{A}_{6} \lb{HGHG2bis} \\
&& + \int \frac{d^3x}{2 \kappa} \gamma_0^2\barn (\delta N_2 -\delta N_1)  ( \delta K^d_d )\mathcal{A}_7 \lb{HGHG5} \\
&& + \int \frac{d^3x}{2 \kappa} \gamma_0^2\frac{\barn}{\barp} (\delta N_2 -\delta N_1)( \partial_c \partial^d \delta E^c_d)  \mathcal{A}_8 \lb{HGHG6} \\
&& + \int \frac{d^3x}{4 \kappa} \gamma_0^2\frac{\barn}{\barp}  (\delta N_2 -\delta N_1) ( \delta E^d_d ) \mathcal{A}_9   \lb{HGHG7} \\
&& + \int \frac{d^3x}{2\kappa}   \gamma_0  \frac{\barn}{\barp}  (\delta N_2 -\delta N_1)(\partial_c\partial^c\delta E_d^d)\mathcal{A}_{10} \lb{HGHG71}\\ 
&& + \int \frac{d^3x}{\kappa} \gamma_0 \frac{\barn}{\barp} \Delta(\delta N_2 -\delta N_1)(\partial_c \partial^d \delta E^c_d)\mathcal{A}_{11} \lb{HGHG72},
\end{eqnarray}

where $\Delta$ corresponds to the Laplacian (that is $\Delta X = \partial_c \partial^c X$). \\ From this expression one can notice that:
\begin{itemize}
\item Eq. (\ff{HGHGD}) is proportional to the gravitational diffeomorphism constraint, as it should be classically. However, the factor in front of the constraint now depends on the counterterms $\alpha_3$ and $\alpha_4$ . 
\item We get a new set of anomalies that must vanish.
\end{itemize}

Two of the anomalies are
\begin{eqnarray}
\mathcal{A}_{10}&=&-(1+\alpha_3)\left[(\bark^2+\alpha_2)\gamma_2+6\bark^2\barp\gamma_8\right],\\
\mathcal{A}_{11}&=&(1+\alpha_3)^2\gamma_2.
\end{eqnarray}
Together with Eq (\ref{g1}), canceling these anomalies leads to
\begin{equation}\lb{g2}
\gamma_1=\gamma_2=\gamma_8=0, 
\end{equation}
which consequently allows one to simplify the remaining anomalies such that: 
\begin{equation}
 \mathcal{A}_{6} =  (1 + \alpha_3) ( \alpha_5- \alpha_4),
\end{equation}
\begin{eqnarray}\lb{A5}
 \mathcal{A}_{7} &=& - \bark^2 \alpha_4 + 3 \bark^2 \alpha_5 + \alpha_2 (2 - \alpha_4+ 3\alpha_5) \nonumber \\
&&  - 4  \alpha_6  (\bark +\alpha_1)+ 4 \bark  \Lambda_{\gamma} (\bark + 2 \alpha_1)  \nonumber \\
&& - 2 \bark^2 \frac{\partial \alpha_1}{\partial \bark } \left( 1+2 \Lambda_{\gamma}\right)  \nonumber \\
&&  +4 \bark \barp   \frac{\partial \alpha_1}{\partial \barp } \left( 2+ \frac{\bark}{\gamma_0} \frac{\partial \gamma_0}{\partial \bark}\right) ,
\end{eqnarray}
\begin{eqnarray}\lb{A6}
 \mathcal{A}_{8} &=& -2 (\bark \alpha_4 +\alpha_6) + 2 \bark \alpha_9 +  2 \alpha_1 (1 + \alpha_9 )  \nonumber \\ 
&& - 4\bark \Lambda_{\gamma}  - 2 \alpha_3 (\bark + \bark \alpha_4 + \alpha_6 + 2 \bark \Lambda_{\gamma}) \nonumber \\
&&+\bark^2 \frac{\partial \alpha_3}{\partial \bark} (1+2\Lambda_{\gamma})  \nonumber \\
&& - 2\bark \barp \frac{\partial \alpha_3}{\partial \barp} \left( 2+ \frac{\bark}{\gamma_0} \frac{\partial \gamma_0}{\partial \bark}\right) \nonumber \\
&& +  2 \frac{\bark^2}{\gamma_0} \frac{\partial \gamma_0}{\partial \bark} (1+\alpha_3)
\end{eqnarray}
and 
\begin{eqnarray}\lb{A7}
 \mathcal{A}_{9} &=& 2 (\bark^2 + \alpha_2) \alpha_6 + 2 \alpha_1 (\bark^2 - 2 \alpha_7 + 3 \alpha_8 ) \nonumber \\
&& + 2 \bark (3 \alpha_8-2\alpha_7+2 \alpha_2 \Lambda_{\gamma}  )  \nonumber \\
&&-\bark^2  \frac{\partial \alpha_2}{\partial \bark} (1+2 \Lambda_{\gamma}) \nonumber \\
&& + 2\bark \barp \frac{\partial \alpha_2}{\partial \barp} \left( 2+ \frac{\bark}{\gamma_0} \frac{\partial \gamma_0}{\partial \bark}\right) \nonumber \\
&& -  2 \frac{\bark^2}{\gamma_0} \frac{\partial \gamma_0}{\partial \bark} (\bark^2+\alpha_2).
\end{eqnarray}

\subsubsection{$\{ H_G[N], D_{m}[N^a]\}$}

This Poisson bracket vanishes:
\begin{equation}
\{ H_G[N], D_{m}[N^a] \} = 0,
\end{equation}
and as a result of this
\begin{equation}
\{ H_G[N], D_{tot}[N^a] \} = \{ H_G[N], D_{G}[N^a] \}.
\end{equation}

\subsubsection{$\{ H_m[N], D_{tot}[N^a] \} $}

This Poisson bracket is given by:

\begin{eqnarray}
&& \{ H_m[N], D_{tot}[N^a] \} = - H_m [\delta N^c \partial_c \delta N] \lb{HMDTH} \\
&&+\int d^3x (  \delta N  \partial_c \delta N^c)\left( V \barp^{\frac{3}{2}} \right) \mathcal{A}_{12} \lb{HMDT8}\\
&&+\int d^3x (  \delta N  \partial_c \delta N^c)\left( \frac{\nu_0 \barpi^2}{2  \barp^{\frac{3}{2}}}  \right) \mathcal{A}_{13} \lb{HMDT9}\\
&& + \int d^3x \barn (\partial_c \delta N^c) \left( V' \barp^{\frac{3}{2}}  \delta \varphi \right)\mathcal{A}_{14}   \lb{HMDT10} \\
&&+ \int d^3x \barn (\partial_c \delta N^c) \left(\frac{\nu_0 \barpi}{ \barp^{\frac{3}{2}}} \delta \pi \right)\mathcal{A}_{15}  \lb{HMDT11}\\
&&-\int d^3x  \barn  ( \delta N^c \partial_d \delta E^d_c )\left( \frac{1}{2} V \sqrt{\barp} \right) \mathcal{A}_{16} \lb{HMDT12}\\
&&+\int d^3x  \barn  ( \delta N^c \partial_d \delta E^d_c )\left( \frac{\nu_0 \barpi^2}{12  \barp^{\frac{5}{2}}}  \right) \mathcal{A}_{17} \lb{HMDT13}\\
&&-\int d^3x  \barn  ( \delta N^c \partial_c \delta E^d_d )\left( \frac{1}{2} V \sqrt{\barp}  \right) \mathcal{A}_{18} \lb{HMDT14}\\
&&+\int d^3x  \barn  ( \delta N^c \partial_c \delta E^d_d )\left( \frac{\nu_0 \barpi^2}{4  \barp^{\frac{5}{2}}}  \right) \mathcal{A}_{19} \lb{HMDT15} \\
&&+\int d^3x  \barn (\delta N^c \partial_d \delta K^d_c)\left( \frac{\barpi^2}{6\barp^{\frac{3}{2}}} \right) \mathcal{A}_{20}  \lb{HMDT151}\\
&&+\int d^3x  \barn (\delta N^c \partial_c \delta K^d_d)\left( \frac{\barpi^2}{6\barp^{\frac{3}{2}}} \right) \mathcal{A}_{21} . \lb{HMDT152}
\end{eqnarray}

In this case:
 
\begin{itemize}
\item As for Eq. (\ff{HGDGH}), Eq. (\ff{HMDTH}) is what is expected classically. Because only the total algebra is physically relevant, it should be noticed that $\{H_{tot},D_{tot}\}$ will also be proportional to $H_{tot}$, as expected.  
\item Moreover, one can see that, after comparing with $\{ H_G[N], D_G[N^c] \} $, there is no term proportional to the matter diffeomorphism constraint, in a similar way as the one in Eq. (\ref{HGDGD}) for the gravitational sector. Therefore, one can conclude, as  previously stated, that Eq. (\ref{HGDGD}) corresponds also to an anomaly, $\mathcal{A}_{22}$, whose expression is given in the following. 
\end{itemize}

The new anomalies are then:
\begin{equation} \lb{A9}
\mathcal{A}_{12} = \beta_{4},
\end{equation}
\begin{equation}
\lb{A8}
\mathcal{A}_{13}=  2\beta_1 - \beta_2 + 2 \bark \frac{\nu_2}{\nu_0},
\end{equation}
\begin{equation}
\lb{A10}
\mathcal{A}_{14} = \beta_{11},
\end{equation}
\begin{equation}
\lb{A11}
\mathcal{A}_{15} =\beta_5 - \beta_6 + \bark \frac{\nu_2}{\nu_0} (1+\beta_1) ,
\end{equation}
\begin{equation}
\lb{A12}
\mathcal{A}_{16} =\beta_{13},
\end{equation}
\begin{equation}
\lb{A13}
\mathcal{A}_{17} =3 \beta_8 + 2 \Lambda_{\nu} -6\bark\barp\frac{\nu_7}{\nu_0} ,
\end{equation}
\begin{equation}
\lb{A14}
\mathcal{A}_{18} =\beta_{12}-\beta_{13}
\end{equation}
\begin{equation}
\lb{A15}
\mathcal{A}_{19} =2\beta_6 -\beta_7 -\beta_8 + \bark\frac{\nu_2}{\nu_0}(1+\beta_2)-2\bark\barp\frac{\nu_8}{\nu_0}, \quad\quad
\end{equation}
\begin{equation}
\lb{A151}
\mathcal{A}_{20} =-6\bark\nu_5+3\barp\nu_7-\frac{\partial\nu_0}{\partial\bark},
\end{equation}
\begin{eqnarray}
\lb{A152}
\mathcal{A}_{21} &=&3(-1-2\beta_1+\beta_3)\nu_2-6\bark\nu_6\nonumber\\
&&-3\barp(\nu_7+2\nu_8) + \frac{\partial\nu_0}{\partial\bark},
\end{eqnarray}
\begin{equation}
\mathcal{A}_{22} =\bark \alpha_4 +\alpha_6 - \frac{\bark^2}{\gamma_0} \frac{\partial \gamma_0}{\partial \bark} .
\end{equation}

\subsubsection{$\{ H_m[N_1], H_m[N_2]\} $}

This Poisson bracket is given by:

\begin{eqnarray}
&&\{ H_m[N_1], H_m[N_2]\} = \nonumber \\
&&  \nu_0^2 (1+\beta_1)(1+\beta_9)  D_m\left[ \frac{\barn}{\barp} \partial^a (\delta N_2-\delta N_1)\right] \lb{HMHMD} \\
&& + \int d^3x \barn (\delta N_2 - \delta N_1) (\delta \varphi) 
\left( V'' \nu_0 \barpi \right) \mathcal{A}_{23}  \lb{HMHM16}\\
&& + \int d^3x \barn (\delta N_2 - \delta N_1) (\delta \varphi) \left(-  \frac{\kappa}{2 } V V' \barp^2 \right ) \mathcal{A}_{24}\lb{HMHM17} \,\,\, \,\,\, \,\,\, \,\,\, \,\,\, \,\,\, \,\,\, \,\,\, \,\,\, \,\,\,\\
&& + \int d^3x \barn (\delta N_2 - \delta N_1) (\delta \varphi) \left(\frac{\kappa }{12 p} \nu_0 V' \barpi^2  \right) \mathcal{A}_{25}\lb{HMHM18} \,\,\, \,\,\,\\
&& + \int d^3x \barn (\delta N_2 - \delta N_1) (\delta \pi ) \left( \nu_0 V'  \right)  \mathcal{A}_{26}\lb{HMHM19}\\
&& + \int d^3x \barn (\delta N_2 - \delta N_1) (\delta \pi ) \left( \frac{\kappa}{2\barp}\nu_0 V \bar{\pi}    \right)  \mathcal{A}_{27}\lb{HMHM191}\\
&& + \int d^3x \barn (\delta N_2 - \delta N_1) (\delta \pi ) \left( - \kappa \frac{\nu_0}{12 \barp^4} \barpi^3 \right)  \mathcal{A}_{28}\lb{HMHM20}\\
&& + \int d^3x \barn (\delta N_2 - \delta N_1)(\delta E^d_d) \left(\frac{\kappa}{4}V^2 \barp  \right)\mathcal{A}_{29} \lb{HMHM21}\\
&& + \int d^3x \barn (\delta N_2 - \delta N_1)(\delta E^d_d) \left(\frac{\nu_0}{2 \barp }V \barpi \right)\mathcal{A}_{30} \lb{HMHM22} \\
&& + \int d^3x \barn (\delta N_2 - \delta N_1)(\delta E^d_d) \left(\frac{\kappa\nu_0}{16 \barp^2 }m^2 \barpi^2 \bar{\varphi}^2 \right)\mathcal{A}_{31} \lb{HMHM221} \\
&& + \int d^3x \barn (\delta N_2 - \delta N_1)(\delta E^d_d) \left(\frac{\kappa\nu_0^2}{48 \barp^5}  \barpi^4 \right)\mathcal{A}_{32} \lb{HMHM23} \\
&& + \int d^3x \barn (\delta N_2 - \delta N_1)(\delta K^d_d) \left(\frac{\kappa}{3 \barp}V \barpi^2  \right)\mathcal{A}_{33}  \lb{HMHM231} \\
&& + \int d^3x \barn (\delta N_2 - \delta N_1)(\delta K^d_d) \left(V' \barpi  \right)\mathcal{A}_{34}  \lb{HMHM232} \\
&& + \int d^3x \barn (\delta N_2 - \delta N_1)(\delta K^d_d) \left(\frac{\kappa}{24 \barp^4} \barpi^4 \right)\mathcal{A}_{35}  \lb{HMHM233}.
\end{eqnarray}

It should be noticed that:

\begin{itemize}
\item 
Eq. (\ff{HMHMD}) is equivalent to Eq. (\ff{HGHGD}) but with a different factor in front. In order to get them compatible, we can re-express Eq. (\ff{HMHMD}) as
\begin{eqnarray}
 && \nu_0^2(1+\beta_1)(1+\beta_9)  D_m\nonumber \\
&=&  \gamma_0^2 (1+\alpha_3)(1+\alpha_4)  D_m+ \mathcal{A}_{36} \,\,D_m,
\end{eqnarray}
where we have introduced a new anomaly $\mathcal{A}_{36} $.
\end{itemize}

The anomalies from this Poisson bracket are then given by :
\begin{eqnarray}
\lb{A16}
\mathcal{A}_{23} &=& -\beta_3 + \beta_{10} + \beta_1 + \beta_1 \beta_{10},\\ \nonumber\\
\lb{A17}
\mathcal{A}_{24} &=& \frac{\partial \beta_3}{\partial \bark},
\end{eqnarray}
\begin{eqnarray}
\lb{A18}
\mathcal{A}_{25} &=& (3-2\Lambda_{\nu})  \frac{\partial \beta_3}{\partial \bark}\nonumber\\
&&+\frac{1}{\nu_0}\frac{\partial\nu_0}{\partial\bark}\left(3(1+\beta_3)+2\barp\frac{\partial\beta_3}{\partial\barp}\right)\nonumber\\ 
&&-9\frac{\nu_2}{\nu_0}(1+\beta_{11}),\\ \nonumber\\
\lb{A19}
\mathcal{A}_{26} &=&\beta_1-\beta_5-\beta_3-\beta_3 \beta_5,\\ \nonumber\\
\lb{A191}
\mathcal{A}_{27} &=&-\frac{\partial\beta_1}{\partial\bark}+3\frac{\nu_2}{\nu_0}(1+\beta_1)(1+\beta_4)\nonumber\\
&&-(1+\beta_1)\frac{1}{\nu_0}\frac{\partial\nu_0}{\partial\bark}\\ \nonumber\\
\lb{A20}
\mathcal{A}_{28} &=&(-3 + 2 \Lambda_{\nu})  \frac{\partial \beta_1}{\partial \bark}
-\frac{2p}{\nu_0}\frac{\partial\nu_0}{\partial\bark}\frac{\partial\beta_1}{\partial\barp}\nonumber\\ 
&&+9\frac{\nu_2}{\nu_0}(\beta_1+\beta_2+\beta_1\beta_2-\beta_6),\\ \nonumber\\
\lb{A21}
\mathcal{A}_{29} &=&-\frac{\partial \beta_4}{\partial \bark},\\ \nonumber\\
\lb{A22}
\mathcal{A}_{30} &=& \beta_1-\beta_2+\beta_3-\beta_4 \nonumber\\
&&+(1+\beta_3)\beta_6+(1+\beta_1)\beta_{11},
\end{eqnarray}
\begin{eqnarray}
\lb{A22.1}
\mathcal{A}_{31} &=&
\frac{\partial\beta_2}{\partial \bark}+
\left(1+\frac{2p}{3\nu_0}\frac{\partial\nu_0}{\partial\barp}\right)\frac{\partial\beta_4}{\partial\bark}\nonumber\\
&&+\frac{1}{3\nu_0}\frac{\partial\nu_0}{\partial\bark}\left(4+3\beta_2+\beta_4+2\barp\frac{\partial\beta_4}{\partial\barp}\right)\nonumber\\
&&+\frac{\nu_2}{\nu_0}(-4-3\beta_2-3\beta_4-3\beta_2\beta_4-3\beta_{12}+2\beta_{13})\nonumber\\
&&+\frac{2\barp}{\nu_0}(\nu_7+3\nu_8)(1+\beta_4),
\end{eqnarray}
\begin{eqnarray}\lb{A23}
\mathcal{A}_{32} &=&
(3-2\Lambda_{\nu}) \frac{\partial \beta_2}{\partial \bark}\nonumber\\
&&+\frac{1}{2\nu_0}\frac{\partial\nu_0}{\partial\bark}\left(1+\beta_2+\barp\frac{\partial\beta_2}{\partial\barp}\right)\nonumber\\
&&+\frac{3\nu_2}{\nu_0}(-2+6\beta_2+3\beta_2^2-3\beta_7-2\beta_8)\nonumber\\
&&-\frac{6\barp}{\nu_0}(\nu_7+3\nu_8)(1+\beta_2),
\end{eqnarray}

\begin{eqnarray}
\mathcal{A}_{33}  &=& (2\nu_5+6\nu_6)(1+\beta_4)-\frac{\partial\nu_2}{\partial\bark},\\ \nonumber\\
\mathcal{A}_{34}  &=& -\nu_2(\beta_1+\beta_2+\beta_1\beta_2),\\ \nonumber\\
\mathcal{A}_{35}  &=& 9\nu_2^2(1+\beta_1)-6\nu_0(\nu_5+3\nu_6)(1+\beta_2)\nonumber\\
&&-6\nu_2\barp(\nu_7+3\nu_8)+\left(-3\nu_2+2\barp\frac{\partial\nu_2}{\partial\barp}\right)\frac{\partial\nu_0}{\partial\bark}\nonumber\\
&&+(3-2\Lambda_\nu)\nu_0\frac{\partial\nu_2}{\partial\bark},
\end{eqnarray}
and
\begin{equation}\lb{A24}
\mathcal{A}_{36}  =  \gamma_0^2 (1+\alpha_3)(1+\alpha_4)-  \nu_0^2(1+\beta_1)(1+\beta_9).
\end{equation}

\subsubsection{$\{ H_G[N_1], H_m[N_2]\} - (N_1 \leftrightarrow N_2) $}
This Poisson bracket reads as:
\begin{eqnarray}
&& \{ H_G[N_1], H_m[N_2]\} - (N_1 \leftrightarrow N_2)= \nonumber \\
&& \int d^3x \barn (\delta N_2 - \delta N_1 ) (\delta \pi) \left( -\frac{\gamma_0}{2  \barp^2} \barpi \nu_0 \right) \mathcal{A}_{37}  \lb{diff25}\\
&& + \int d^3x \barn \Delta(\delta N_2 - \delta N_1 ) (\delta \pi) \left( \frac{\gamma_0}{  \barp^2} \barpi  \right) \mathcal{A}_{38}  \lb{diff251}\\
&& + \int d^3x \barn (\delta N_2 - \delta N_1 ) (\delta \varphi)\left(\frac{\gamma_0}{2 } V' \barp  \right) \mathcal{A}_{39}  \lb{diff26}\\
&& + \int d^3x \barn (\delta N_2 - \delta N_1 ) (\delta K^d_d)\left(\frac{\gamma_0}{12  \barp^2} \nu_0 \barpi^2\right)\mathcal{A}_{40}  \lb{diff27}\\
&& + \int d^3x \barn (\delta N_2 - \delta N_1 ) (\delta K^d_d)\left(\frac{\gamma_0}{2}\barp V\right)\mathcal{A}_{41} \lb{diff28}\\
&& + \int d^3x \barn (\delta N_2 - \delta N_1 ) (\delta E^d_d)\left( \frac{\gamma_0}{12  \barp^3}\frac{\nu_0}{2} \barpi^2 \right) \mathcal{A}_{42} \qquad  \,\,\,   \,\,\, \,\,\, \,\,\, \,\,\,\lb{diff29}\\
&& + \int d^3x \barn (\delta N_2 - \delta N_1 ) (\delta E^d_d) \left(\frac{\gamma_0}{4}  V \right) \mathcal{A}_{43} \lb{diff30}\\
&& +\int d^3x \barn (\delta N_2 - \delta N_1 ) (\partial_c \partial^d \delta E^c_d) \left(\frac{\gamma_0}{12\barp^3}\barpi^2\right)  \mathcal{A}_{44} \lb{diff311}\\
&& +\int d^3x \barn (\delta N_2 - \delta N_1 ) (\partial_c \partial^d \delta E^c_d) \left(\frac{\gamma_0}{2}V\right)  \mathcal{A}_{45} \lb{diff312}\\
&& -\int d^3x \barn (\delta N_2 - \delta N_1 ) (\Delta \delta E^d_d) \left(\frac{\gamma_0}{4\barp^3}\barpi^2\right)  \mathcal{A}_{46} \lb{diff32}\\
&& +\int d^3x \barn (\delta N_2 - \delta N_1 ) (\partial_c\partial^d \delta K^c_d) \left(\frac{\gamma_0}{\barp^2}\barpi^2\right)  \mathcal{A}_{47} \lb{diff33}\\
&& +\int d^3x \barn (\delta N_2 - \delta N_1 ) (\Delta \delta K^d_d) \left(\frac{\gamma_0}{\barp^2}\barpi^2\right)  \mathcal{A}_{48}. \lb{diff34}
\end{eqnarray}

After using Eq. (\ref{g2}), we find that
\begin{eqnarray}
\mathcal{A}_{38}  &=& (1+\alpha_3)(1+\beta_1)\nu_2,\\
\mathcal{A}_{46} &=& (1+\alpha_3)(1+\beta_2)\nu_2,\\
\mathcal{A}_{47} &=& (1+\alpha_3)\nu_5,\\
\mathcal{A}_{48} &=& (1+\alpha_3)\nu_6.
\end{eqnarray}
Demanding that the above anomalies do vanish, while respecting the classical limit,  leads to:
\begin{equation}\lb{n2}
\nu_2=\nu_5=\nu_6=0.
\end{equation}

Using Eqs. (\ref{g2}) and (\ref{n2}) leads to the remaining anomalies:
\begin{eqnarray}\lb{A25}
\mathcal{A}_{37}   &=&
6 \alpha_1 -6\bark \beta_1 +6 (\bark + \alpha_1)\beta_6 \nonumber \\
&& - \bark^2(1+2\Lambda_{\gamma}) \frac{\partial \beta_1}{\partial \bark} + 2 \bark \barp  \frac{\partial \beta_1}{\partial \barp} \left(2 + \frac{\bark}{\gamma_0} \frac{\partial \gamma_0}{\partial \bark} \right) \nonumber \\
&& + (1+\beta_1)\left(4 \bark \Lambda_\nu - \frac{\bark^2}{\gamma_0} (3-2 \Lambda_\nu) \frac{\partial \gamma_0}{\partial \bark} \right) \nonumber \\
&&-\frac{\bark^2}{\nu_0}\frac{\partial\nu_0}{\partial\bark}(1+\beta_1)(1+2\Lambda_\gamma),
\end{eqnarray}
\begin{eqnarray}\lb{A26}
\mathcal{A}_{39}  &=&  6 (\bark + \alpha_1)  (1+\beta_{11})+ \bark^2 (1 + 2 \Lambda_\gamma) \frac{\partial \beta_3}{\partial \bark} \nonumber \\
&& - \bark \left( 2 + \frac{\bark}{\gamma_0} \frac{\partial \gamma_0}{\partial \bark} \right)\left( 3 (1+\beta_3) +2 \barp \frac{\partial \beta_3}{\partial \barp}  \right), \qquad. 
\end{eqnarray}
\begin{eqnarray}\lb{A27}
&& \mathcal{A}_{40} =  6\beta_2 + (9\alpha_5-3\alpha_4)(1+\beta_2)+4\Lambda_\nu \nonumber \\
&& +\left(
\frac{(4\Lambda_\nu-6)}{\gamma_0}\frac{\partial \gamma_0}{\partial \bark}+
\frac{12\barp}{\nu_0}(\nu_7+3\nu_8)-\frac{2+4\Lambda_\gamma}{\nu_0}
\right)(\bark+\alpha_1) 
\nonumber \\
&& +(4\Lambda_\nu-6)\frac{\partial \alpha_1}{\partial \bark}-\frac{4\barp}{\nu_0}\frac{\partial \nu_0}{\partial \bark}\frac{\partial \alpha_1}{\partial \barp},
\end{eqnarray}
\begin{eqnarray}\lb{A28}
\mathcal{A}_{41}& =& -2\beta_4 + (\alpha_4-3\alpha_5)(1+\beta_4) \nonumber \\
&& + 2\frac{\partial \alpha_1}{\partial \bark} + \frac{2}{\gamma_0}(\bark + \alpha_1)\frac{\partial \gamma_0}{\partial \bark},
\end{eqnarray}
\begin{eqnarray}\lb{A29}
 \mathcal{A}_{42} &=& - 3 \bark^2 \frac{\partial \beta_2}{\partial \bark} ( 1 + 2 \Lambda_\gamma) \nonumber \\
&&+6 (\bark +\alpha_1)(5 + 3 \beta_7 + 2 \beta_8) +6 ( \bark + \alpha_6)(1+ \beta_2) \nonumber \\
&& + 3\bark \left(2+\frac{\bark}{\gamma_0}\frac{\partial \gamma_0}{\partial\bark} \right)\left((1+\beta_2)(-5 + 2 \Lambda_\nu) +2 \barp \frac{\partial  \beta_2 }{\partial \barp} \right) \nonumber \\
&& + \frac{(-3+2 \Lambda_\nu)}{\gamma_0} \frac{\partial }{\partial \bark } \left( \gamma_0 (\alpha_2 +\bark^2) \right)  \nonumber \\
&& - \frac{6\barp}{\nu_0}(\nu_7+3\nu_8)(\bark^3+\alpha_2) \nonumber \\
&& - \frac{1}{\nu_0}\frac{\partial  \nu_0 }{\partial \bark}\left(2k^2-\alpha_2+3k^2\beta_2+2p\frac{\partial  \alpha_2 }{\partial \barp}\right) \nonumber \\
&& - \frac{2\Lambda_\gamma}{\nu_0}\frac{\partial  \nu_0 }{\partial \bark}\left(4k^2+\alpha_2+3k^2\beta_2\right),
\end{eqnarray}
\begin{eqnarray}\lb{A30}
\mathcal{A}_{43} &=&- 2 (\bark + \alpha_6)(1+\beta_4) + (\bark +\alpha_1)(-4 \beta_{13} + 6 \beta_{12}+2) \nonumber \\
& & - \bark \left(2+\frac{\bark}{\gamma_0} \frac{\partial  \gamma_0}{\partial \bark} \right) \left( \beta_4 + 2 \barp\frac{\partial \beta_4}{\partial \barp} \right) \nonumber \\
&&+ \bark^2 \frac{\partial \beta_4}{\partial \bark} (1+2 \Lambda_\gamma) + \frac{1}{\gamma_0}\frac{\partial }{\partial \bark } \left( \gamma_0 \alpha_2  \right),
\end{eqnarray}
\begin{eqnarray}\lb{A31}
\mathcal{A}_{44}&=&\frac{\nu_0(3-2\Lambda_\nu)}{\gamma_0}\frac{\partial}{\partial\bark}\gamma_0(1+\alpha_3) \nonumber\\
&&+\frac{\partial\nu_0}{\partial\bark}\left((2\Lambda_\gamma-1)(1+\alpha_3)+2p\frac{\partial\alpha_3}{\partial\barp}\right)\nonumber\\
&&+6p(\nu_7+\nu_8)(1+\alpha_3),
\end{eqnarray}
and
\begin{equation}\lb{A312}
\mathcal{A}_{45}=-\frac{1}{\gamma_0}\frac{\partial}{\partial\bark}\gamma_0(1+\alpha_3).
\end{equation}

\subsubsection{$\{D_{tot} [N^a], D_{tot}[N^b] \}$}
Finally,
\begin{equation}
\{D_{tot} [N^a], D_{tot}[N^b] \} = 0.
\end{equation}

\subsection{Solution after canceling the anomalies}

After solving the anomalies, it can be found that:
\begin{eqnarray}
&&\gamma_1=\gamma_2=\gamma_8=0, \lb{gamma} \\
&&\nu_2=\nu_3=\nu_5=\nu_6=\nu_7=\nu_8=0. \lb{nu}
\end{eqnarray}
As pointed out before, the remaining $\gamma_i$, $\nu_i$, $i\neq0$, can safely be absorbed into the counterterms $\alpha_i$ and $\beta_i$. This leads to an interesting situation: only the corrections expressed with the homogeneous background corrections will affect the algebra and therefore the equations of motion. In other words, the whole effect of the corrections in this work is obtained only by the zeroth order. Nevertheless, this may not be true anymore when the higher order derivatives will be taken into account.

We also find that
\begin{equation}\lb{nuk}
\frac{\partial\nu_0}{\partial\bark}=0,
\end{equation}
that is, $\nu$ has to  depend only on $\barp$.

In order to clarify the equations in the following, we will use the convention 
\begin{eqnarray} 
\Gamma_{\bark} &=& \frac{\bark}{\gamma_0} \frac{\partial \gamma_0 }{\partial \bark}, \\
\Gamma_{\bark,2}&=& \frac{\bark^2 }{\gamma_0}\frac{\partial^2 \gamma_0 }{\partial \bark^2},\\
\Sigma_{\barp}&=&\frac{9-9\Lambda_\nu+2\Lambda_\nu^2+2p\frac{d\Lambda_\nu}{d\barp}}{2\nu_0(3-\Lambda_\nu)^2}. \lb{defSigma}
\end{eqnarray}
It should be noticed that, because of Eq. (\ref{nuk}), $\Sigma_{\barp}$ only depends on $\barp$. In the classical limit $\Gamma_{\bark},\Gamma_{\bark,2}\to0$ and $\Sigma_{\barp}\to\frac{1}{2}$.
\\ After having solved all the previous anomalies without any ambiguity, the 'final' expressions for the counterterms are now given by:  
\begin{eqnarray}
&&\alpha_1 = \bark \Big( -1+ \left(2 + \Gamma_{\bark} \right) \Sigma_{\barp} \Big) ,\lb{alpha1} \\
&& \alpha_2 = - 2 \bark^2 \Big( -1+ \left(2 + \Gamma_{\bark}\right) \Sigma_{\barp}  \Big) , \\
&& \alpha_3 = -1+\frac{f_1[\barp]}{\gamma_0}, \\
&& \alpha_4 = -1+\left(2 + 4 \Gamma_{\bark} + \Gamma_{\bark,2} \right)  \Sigma_{\barp} , \\
&& \alpha_5 = -1+\left(2 + 4 \Gamma_{\bark} + \Gamma_{\bark,2} \right)  \Sigma_{\barp} ,\\
&& \alpha_6 =\bark \Big(1 + \Gamma_{\bark}-\left(2 + 4 \Gamma_{\bark} + \Gamma_{\bark,2} \right)  \Sigma_{\barp}  \Big),\\
&& \alpha_7 = 2 \bark^2 \Big(1- \Lambda_\gamma + \Gamma_{\bark} -\left(2 + 4 \Gamma_{\bark} + \Gamma_{\bark,2} \right)  \Sigma_{\barp}  \Big),\quad\quad\quad \\ 
&& \alpha_8 =  2 \bark^2 \Big(1- \Lambda_\gamma + \Gamma_{\bark} -\left(2 + 4 \Gamma_{\bark} + \Gamma_{\bark,2} \right)  \Sigma_{\barp}  \Big),\\ 
&&\alpha_9 = -1 + \frac{1}{2 \gamma_0 \Sigma_{\barp}} \left(f_1[\barp] + 2 \barp \frac{df_1[\barp]}{d\barp} \right),
\end{eqnarray}
and 
\begin{eqnarray}
&& \beta_1 = -\frac{\Lambda_{\nu}}{3}, \\
&& \beta_2 = -2\frac{\Lambda_{\nu}}{3}\\
&& \beta_3 =-1+\frac{3}{\nu_0(3-\Lambda_\nu)}\\
&&\beta_4=0,\\
&&\beta_5 =  -1 + \frac{\nu_0(3-\Lambda_\nu)^2}{9},\\
&&\beta_6 =  -1 + \frac{\nu_0(3-\Lambda_\nu)^2}{9},\\
&&\beta_7 = \frac{2(3-\Lambda_\nu)\big(3(\nu_0-1)-\Lambda_\nu\nu_0\big)}{9},\\
&&\beta_8 = -2 \frac{\Lambda_{\nu}}{3}, \\
&&\beta_9 = -1 + \frac{3 \gamma_0 \left(2 + 4 \Gamma_{\bark} + \Gamma_{\bark,2} \right) \Sigma_{\barp}}{\nu_0^2 (3-\Lambda_{\nu})} f_1[\barp] ,\\
&&\beta_{10} =-1+\frac{9}{\nu_0(3-\Lambda_\nu)^2}, \\ 
&&\beta_{11} = 0, \\
&&\beta_{12} = 0, \\
&&\beta_{13} = 0, \lb{beta13}
\end{eqnarray}
where $f_1[\barp]$ is an unknown function of $\barp$. In order for all the counterterms to vanish at the classical limit, we require that $f_1[\barp]$ tends towards $1$ and its derivatives towards 0 in this limit.
\\

Some preliminary comments can be made at this stage:
\begin{itemize}

\item[$\bullet$] Firstly, if we consider the different total Poisson brackets, we have now:
\begin{eqnarray}
\left\{ D_{tot}[N^c_1],D_{tot} [N^d_2] \right\} &=& 0, \\
\left\{ H_{tot}[N],D_{tot}[N^c] \right\} &=& - H_{tot}[\delta N^c \partial_c \delta N], \\
\left\{ H_{tot}[N_1],H_{tot}[N_2] \right\} &=&  \Omega D_{tot} \!\! \left[ \!\!  \frac{\bar{N}}{\bar{p}} \partial^c(\delta N_2 - \delta N_1)\right], \,\,\, \,\,\, \,\,\,\,\,\,\, \,\,\,
\label{HtotHtot}
\end{eqnarray}
where $\Omega$ corresponds to the structure function of the modified algebra, given theoretically by 
\begin{eqnarray}
\Omega &=&\gamma_0^2 (1+\alpha_3)(1+\alpha_4)\\
 &=& \nu_0^2 (1+\beta_1)(1+\beta_9). \lb{beta9beta1}
\end{eqnarray}
After using the solutions for the counterterms, we find that in the case of the inverse-volume corrections, $\Omega$ can be expressed as 
\begin{eqnarray}\lb{omega}
\Omega &=& \gamma_0\left(2 + 4 \Gamma_{\bark} + \Gamma_{\bark,2} \right)\Sigma_{\barp}f_1[\barp]\nonumber \\
&=& \left(\frac{\partial^2}{\partial\bark^2}(\bark^2\gamma_0)\right)\Sigma_{\barp}f_1[\barp].
\label{OmegaIV}
\end{eqnarray}
So far, nothing conclusive can be said about the evolution of the structure function with respect to the evolution of the universe. One would need the full theory. Nevertheless, it can be seen here that, as for the holonomy correction, the general expression of $\Omega$ has a possible dependence on $\bark$ or $\barp$ and it is possible that when both corrections will be taken into account simultaneously, a compatible structure function will emerge from the formalism. This would allow one to study a  cosmological scenario which could be close to, or consistent with, the one derived from  loop quantum gravity. Moreover, in this case, it will be interesting to compare the results with what was predicted in \cite{abhay1}.

\item[$\bullet$] Secondly, after having derived the counterterms in Eqs. (\ref{gamma}) - (\ref{beta13}), one additional equation remains : one can notice that $\mathcal{A}_5$ and $\mathcal{A}_7$ appear to be equivalent up to a factor $-k$:
\begin{equation}\label{prerest}
\mathcal{A}_7 \equiv -\bark  \mathcal{A}_5 = 0,
\end{equation}
and requiring that these two anomalies vanish generates a differential equation relating $\gamma_0$ to $\nu_0$, which after some reformulation, becomes:
\begin{eqnarray}\label{rest}
&&\barp
\frac{\partial}{\partial\barp}\ln\left[\gamma_0(2+\Gamma_k)\Sigma_{\barp}^2\right]
=\nonumber \\ &&=
\bark
\frac{(1-\Lambda_\gamma)}{(2+\Gamma_{\bark})}
\frac{\partial}{\partial\bark}\ln\left[
\frac{1-\Lambda_\gamma}{\gamma_0(2+\Gamma_{\bark})^2}
\right].
\end{eqnarray}

This equation can be regarded as an extra constraint on the expression of the inverse-volume corrections, as it has to be fulfilled in order for the total algebra to be closed. Moreover, in order to solve it, one needs either an additional assumption, or the expression from the full theory. In the next section, we will play with the dependence of $\gamma_0$ as the required extra assumption and see what the approach has to say.

\end{itemize}

\subsection{Solution after further assumptions}
So far in this study the only thing that determines the behavior of the inverse-volume corrections is Eq. (\ref{rest}), which is a result of the closure of the algebra, and the fact that we require appropriate conditions on $\gamma_0$ and $\nu_0$ in the classical limit ($\gamma_0\to1$ and $\nu_0\to1$). These constraints are not enough to fully determine the shape of the inverse-volume corrections, therefore further inputs must come from the full theory. Nevertheless, it can be interesting to make some assumptions and see what they do imply. 
In this section, we will first make some assumptions on the dependence of $\gamma_0$ on $\bark$ and $\barp$ and see what are the consequences on $\Omega$, and then, reverse the logic.

\subsubsection{Case where $\gamma_0 = \gamma_0[\barp]$ only }

In most of the literature, the inverse-volume corrections are assumed to depend only on $\barp$.
In this case, Eq. (\ref{rest}) reduces to 
\begin{equation}
\gamma_0\Sigma_{\barp}^2 = \text{constant}.
\end{equation}
Since $\gamma_0\to1$ and $\Sigma_{\barp}\to\frac{1}{2}$ in the classical limit, we must have
\begin{equation}\lb{sigmap}
\Sigma_{\barp}=\frac{1}{2\sqrt{\gamma_0}},
\end{equation}
which, after using the definition of $\Sigma_{\barp}$ given by Eq. (\ref{defSigma}), leads to a differential equation relating $\gamma_0$ with $\nu_0$ (both depending only on $\barp$) such that 
\begin{equation} \lb{diffequanu0}
9-9\Lambda_\nu+2\Lambda_\nu^2- \frac{\nu_0(3-\Lambda_\nu)^2}{\sqrt{\gamma_0}}+2p\frac{d\Lambda_\nu}{d\barp} =0. 
\end{equation}
This equation is  complicated and requires to know  at least the expression for one correction to be solved. Nevertheless, solving all the anomalies leads to :

\begin{eqnarray}
&&\alpha_1 = \bark \left(-1+\frac{1}{\sqrt{\gamma_0}}\right), \\
&& \alpha_2 = 2 \bark^2 \left(1-\frac{1}{\sqrt{\gamma_0}}\right), \\
&& \alpha_3 = -1+\frac{f_1[\barp]}{\gamma_0}, \\
&& \alpha_4 = -1+\frac{1}{\sqrt{\gamma_0}}, \\
&& \alpha_5 = -1+\frac{1}{\sqrt{\gamma_0}}, \\
&& \alpha_6 =\bark \left(1-\frac{1}{\sqrt{\gamma_0}}\right), \\
&& \alpha_7 = 2 \bark^2 \left( 1-\frac{1}{\sqrt{\gamma_0}} - \Lambda_{\gamma}\right), \\
&& \alpha_8 = 2 \bark^2 \left( 1-\frac{1}{\sqrt{\gamma_0}} - \Lambda_{\gamma}\right), \\
&&\alpha_9 = -1 + \frac{1}{\sqrt{\gamma_0}} \left(f_1[\barp] + 2 \barp \frac{df_1[\barp]}{d\barp} \right),
\end{eqnarray}
and 
\begin{equation}
\beta_9 = -1 + \frac{3\sqrt{\gamma_0}}{\nu_0^2 (3-\Lambda_\nu)} f_1[\barp], 
\end{equation}
the other $\beta_i$ remaining unchanged. It can be also noticed that in this case, Eq. (\ref{omega}) becomes:
\begin{equation}\lb{omegap}
%\Omega= \frac{f_1[p]}{\sqrt{\gamma_0}}.
\Omega= f_1[p] \times \sqrt{\gamma_0}.
\end{equation}
So far, nothing has been assumed on the shape of $f_1[p]$ except that it has to go to 1 at the classical limit. In this case, one could assume correctly that $ f_1[p] = \frac{1}{\sqrt{\gamma_0}} $ such that $\Omega=1$ and  $\alpha_3 = -1+\frac{1}{{\gamma_0}^{\frac{3}{2}}}$. \\
This case is interesting because it shows that even when dealing with some perturbed and quantum-corrected constraints, the algebra remains closed and one recovers the usual spacetime of general relativity, a consequence which was not expected when considering the previous results derived with the holonomy correction alone. 
%In other word, the Hojman-Kuchar-Teitelboim theorem  \cite{Hojman:1976vp} does not apply here. 

\subsubsection{Case where $\gamma_0 = \gamma_0\left[\frac{\bark}{\sqrt{\barp}}\right]$}
After applying corrections, such as the inverse-volume or holonomy corrections, the resulting physical observable (i.e. the Hubble parameter) are generally not independent of a rescaling of the scale factor. This is a problem since rescaling the scale factor is a gauge choice and therefore not physical. But in the $\bar{\mu}$-scheme for holonomy corrections and $\gamma_0=\gamma_0[\frac{\bark}{\sqrt{\barp}}]
$ for the inverse-volume corrections, the physics becomes invariant under this rescaling. This is why we here consider this case: $\gamma_0 = \gamma_0\left[\frac{\bark}{\sqrt{\barp}}\right]$.

In this section, a prime will denote derivative with respect to the argument
$\frac{\bark}{\sqrt{\barp}}$, 
\begin{equation}
\gamma_0':=\frac{d\gamma_0[\frac{\bark}{\sqrt{\barp}}]}{d\frac{\bark}{\sqrt{\barp}}}.
\end{equation}
Interesting consequences arise in this example. Indeed, in this case Eq. (\ref{rest}) reduces to
\begin{equation} \lb{eqSigma05}
\Sigma_{\barp}=\frac{1}{2}.
\end{equation}

The obtained counterterms are:
\begin{eqnarray}
&&\alpha_1=\frac{k^2\gamma_0'}{2\sqrt{\barp}\gamma_0},\\
&&\alpha_2=-\frac{k^3\gamma_0'}{\sqrt{\barp}\gamma_0},\\
&&\alpha_3=-1+\frac{f_1[\barp]}{\gamma_0},\\
&&\alpha_4=\frac{k\left(4\sqrt{\barp}\gamma_0'+k\gamma_0''\right)}{2\barp\gamma_0},\\
&&\alpha_5=\frac{k\left(4\sqrt{\barp}\gamma_0'+k\gamma_0''\right)}{2\barp\gamma_0},\\
&&\alpha_6=\frac{k^2\left(2\sqrt{\barp}\gamma_0'+k\gamma_0''\right)}{2\barp\gamma_0},\\
&&\alpha_7=-\frac{k^3\left(\sqrt{\barp}\gamma_0'+k\gamma_0''\right)}{\barp\gamma_0},\\
&&\alpha_8=-\frac{k^3\left(\sqrt{\barp}\gamma_0'+k\gamma_0''\right)}{\barp\gamma_0},\\
&&\alpha_9=-1+\frac{1}{\gamma_0} \left(f_1[\barp] + 2 \barp \frac{df_1[\barp]}{d\barp} \right).
\end{eqnarray}
%and
%\begin{eqnarray}
%&&\beta_i= \text{\gren{\dots}}
%\end{eqnarray}
%\gren{The $\beta$'s are not as nice looking as the $\alpha$'s in this case. Should I write them anyway?}\\

The expression for the $\beta_i$ terms are much more complicated and are not given here. Nevertheless, the structure function remains simple and becomes
\begin{equation}
\Omega=\left(\gamma_0+2\frac{\bark}{\sqrt{\barp}}\gamma_0'+\frac{1}{2}\left(\frac{\bark}{\sqrt{\barp}}\right)^2\gamma_0''\right)f_1[\barp].
\end{equation}
In this example, $\Omega=\Omega\left[\frac{\bark}{\sqrt{\barp}}\right]$ if and only if $f_1\equiv1$. \\ As a toy model, one can also assume that $f_1\equiv1 \equiv \Omega$ and solve directly the corresponding differential equation : the solution is given by 
\begin{equation} \lb{solIVkp}
\gamma_0 =1+ D_1 \frac{\sqrt{\barp}}{\bark} + D_2 \frac{\barp}{\bark^2},
\end{equation} 
and one can see that it blows up at the classical limit, for any constants $D_i\neq0$. If $D_1=D_2=0$, then the above expression remains finite, however there will be no inverse-volume correction in the gravitational sector, which is not a good solution.\\

Nevertheless, the interesting part comes from Eq. (\ref{eqSigma05}) leading to an equation similar to Eq. (\ref{diffequanu0}), whose solution is given by
\begin{equation}\lb{solnu}
\nu_0=2\frac{\barp^3}{C_1^2}\left(\frac{C_1}{p^{3/2}}-\ln\left[1+\frac{C_1}{p^{3/2}}\right]\right),
\end{equation}
where $C_1$ is an unknown constant. In the calculations, the classical limits have been respected, such that $\lim_{p\to\infty}\nu_0=1$. Moreover, it should be noticed that 
\begin{equation}
\nu_0=\lim_{C_1\to0} 2\frac{\barp^3}{C_1^2}\left(\frac{C_1}{p^{3/2}}-\ln\left[1+\frac{C_1}{p^{3/2}}\right]\right)=1
\end{equation}
is also solution. Surprisingly, when one tries to plot these expressions for a large range of given values of $C_1$, one sees that the evolution of $\nu_0$ corresponds exactly to the evolution expected for an inverse-volume correction, namely an evolution similar to the one given by Eq. (\ref{evolIV}). From this perspective, it can be understood that assuming that the inverse-volume depends also on the connection is consistent and leads to a good behavior, at least for the correction in the matter sector. \\
The results obtained here could suggest that with a quite similar dependence for $\gamma_0$ on $\bark$, it would be possible to have also a good expression for $\nu_0$ and, more importantly, a good expression for $\gamma_0$ with $\Omega=1$ in the classical limit.

\subsubsection{General case where $\Omega=\Omega[\barp]$}
So far, we have seen two examples where specific dependencies of $\gamma_0$ on $\bark $ and $\barp$ were assumed. In this section, we would like to generalize the approach by studying only $\Omega$ and see the consequences of the following assumptions on the inverse-volume correction. 
We have considered one case where if $\gamma_0 =\gamma_0\left[\frac{\bark}{\sqrt{\barp}}\right]$, then $\Omega=\Omega\left[\frac{\bark}{\sqrt{\barp}}\right]$ if and only if $f_1=1$, and another one where $\Omega = 1$. \\
 Moreover, everything following from $\Omega=\Omega[\barp]$ follows for $\Omega= 1 $ and we will now assume only $\Omega=\Omega[\barp]$. Doing this, one can firstly notice that Eq. (\ref{omega}) can be reformulated as
\begin{equation} 
\frac{\partial^2}{\partial\bark^2}(\bark^2\gamma_0)=\frac{\Omega}{\Sigma_{\barp}f_1[\barp]}.
\end{equation}
If we demand that $\Omega$  depends only on $\barp$ and not on $\bark$ as assumed here, we obtain under the condition that $\gamma_0$ and $\Sigma_{\barp}$ fulfills also Eq. (\ref{rest}), the general solution 
\begin{equation}\label{solgamma}
\gamma_0=\frac{\Omega}{2\Sigma_{\barp}f_1[\barp]}+\frac{f_2[p]}{\bark}+\frac{f_3[p]}{\bark^2},
\end{equation}
where $f_2[p]$ and $f_3[p]$ are functions of $\barp$, going to 0 if one assumes $\gamma_0 = \gamma_0[\barp]$. After some assumptions and manipulations, it is easy to see that the solutions previously derived, corresponding to  Eq. (\ref{omegap}) and Eq. (\ref{solIVkp}) can be recovered from Eq. (\ref{solgamma}).  \\
Looking at the expression in Eq. (\ref{solgamma}), one can worry about the behavior of the inverse-volume correction when $\bark \rightarrow 0 $, that is, about the possibility that the correction is diverging at low curvature. The important point is that the correction derived here should not to be considered alone, but rather with the gravitational Hamiltonian constraint density which include a $\bark^2$ in the numerator. As a consequence, one deals finally with this $\bark^2$ coming from the density and the terms $\frac{1}{\bark}$ and $\frac{1}{\bark^2}$ coming from the correction, such that on the one hand, only the Hamiltonian constraints are well defined for every $\bark$, and consequently, on the other hand, the Friedmann equation is indeed modified but finite. \\ Roughly speaking, the Friedmann equation would have in this case an expression similar to  
\begin{equation}\lb{FriedmannIV}
\left(c_1[\barp] \frac{\bark^2}{\barp}+c_2[p] \bark + c_3[p]\right) = \frac{\kappa}{3} \rho,
\end{equation}
where 
\begin{equation}
\rho = \frac{\nu_0}{2} \frac{\barpi^2}{\barp^3} + V(\varphi).
\end{equation}
One can analyze this situation in the simplest toy model where, for instance $c_2[\barp] = 0$ and $H = c_1[\barp] \bark$ : in this case, the former equation leads to  
\begin{equation}\lb{bounceIV}
H^2 = c_1[\barp] \left( \frac{\kappa}{3} \rho[\barp] - c_3[p] \right),
\end{equation}
where a bounce is possible when $\frac{\kappa}{3} \rho[\barp] = c_3[p]$. Of course, this is just a preliminary idea of a possible consequence of the inverse-volume correction, and one should go through the  full equations.\\

Another comment could be made : the goal of effective loop quantum cosmology is to obtain some constraints which would be as close as possible to the one derived from the full theory. Considering the holonomy correction, it can be shown that the constraints used in \cite{WilsonEwing:2012bx}, built somehow as in loop quantum gravity but however considering Lagrange multipliers depending also on the phase space variables, lead after a Taylor expansion around the background, exactly to the one derived in \cite{scalars}, which pioneered the approach depicted here. One could wonder if it is also the case in this study. The answer is not known yet, the inverse-volume correction having not been considered in the first case, as was the holonomy correction. To do so, one could at least notice that during the derivation of the counterterms, it is possible to see from Eq. (\ref{A16}) and Eq. (\ref{beta9beta1}) that $\beta_9$ and $\beta_{10}$ which, with $\beta_3$, are expressed in the potential $\varphi$ part of the matter densities, are proportional to $\frac{1}{1+\beta_1}$, while the others $\beta_i$ (including $\beta_1$) found in the momentum $\pi$ part of the matter densities are mostly proportional to $\frac{1}{1+\beta_3}$. One can then wonder in which way $\beta_9$ and $\beta_{10}$, related to $\beta_3$ in the derivation of the perturbed constraints, are intertwined with $\beta_1$, while the other counterterms, related to $\beta_1$ in the derivation of the perturbed constraints, are related to $\beta_3$. One can therefore wonder how this particular property in  the expressions of the $\beta_i$ counterterms could emerge from the not Taylor-expanded constraint.\\

Finally, it is possible to see here that this result works also for $\Omega= 1$ and one can generalize the comments presented for $\gamma_0 = \gamma_0[\barp]$ : having a perturbed and quantum-corrected constraint whose corrections depend possibly on both the connection and the flux of densitized triad does not necessarily lead to a deformed algebra as suggested in some previous work.

\subsection{Summary of the inverse-volume case}
The situation for the inverse-volume correction can be summarized as follows:

\begin{itemize}
\item It is possible to close the algebra under the assumptions of this work.

\item All the counterterms and higher order inverse-volume corrections can be determined as functions of the  zeroth order inverse-volume corrections and of the unknown functions $f_1[\bark]$ and $\Sigma_{\barp}$. 

\item One can constrain the zeroth order inverse-volume corrections through $\frac{\partial\nu_0}{\partial\bark}=0$ and Eq. (\ref{rest}).

\item All the counterterms have correct classical limits, as required by construction.

\item The case $\gamma_0=\gamma_0\left[\frac{\bark}{\sqrt{\barp}}\right]$ leads to simplified expressions for the counterterms, and also to a consistent solution for $\nu_0$ given by Eq. (\ref{solnu}). 

\item Assuming $\Omega = \Omega[\barp]$ leads to a general expression for $\gamma_0$ given by Eq. (\ref{solgamma}), but also a modified Friedmann  equation, Eq. (\ref{bounceIV}), with a possible bounce (depending on the value of $\barp$).

\item More importantly, it is possible to consider the presence of quantum corrections in  effective loop quantum cosmology without having to deform the algebra, namely $\Omega = 1$ is allowed but with a modified Friedmann equation compatible with a bounce.

\end{itemize}

\section{Holonomy case - reminder} \label{HOL}
In order to understand better what happens when both corrections are taken into account, it is useful to recall briefly the main results obtained with the holonomy corrections. 

In \cite{scalars}, an anomaly-free algebra was constructed for holonomy-corrected constraints. It was then demonstrated that the $\barmu$-scheme is recovered, that is $\omega = -\frac{1}{2}$ in Eq. (\ff{muscheme}), and the function structure was
given by $\Omega = \text{cos}(2 \barmu \gamma \bark)$. The fact that it can change {its} sign {could be} interpreted as a change from an hyperbolic to an elliptic regime around the bounce, or as a change of signature of the metric from $(-,+,+,+)$ in the Lorentzian phase (where $\{H,H\} =+D $ in our convention) to  $(+,+,+,+)$ in the Euclidean phase (where $\{H,H\} =-D $). As $\Omega$ enters directly  the equation of motion of the perturbations, this would lead to a radical change in the resulting spectrum (see \cite{spectrum_linda} and \cite{jakub_cosmo} for an analysis of the spectrum). The surface $\Omega=0$ may also be interpreted as an "asymptotic silence" surface where initial conditions would have to be set \cite{bkl}. The debate is not closed on this point. We simply try here to push the effective approach as far as possible to investigate its consequences.

\subsection{constraints}

The constraints with holonomy corrections read as:
\begin{equation}
D_G[N^c]\!\! = \!\! \int \frac{d^3x}{\kappa} \delta N^c \left[ \barp \partial_c \delta K^d_d -
\partial_d \delta K^d_c - \bark \partial_d \delta E^d_c \right],
\end{equation}
\begin{equation}
H_G[N] =  \frac{1}{2 \kappa} \int d^3x (\barN + \delta N) 
\times \left[ \mathcal{H}^{(0)}_G +\mathcal{H}^{(1)}_G + \mathcal{H}^{(2)}_G \right],
\end{equation}
\begin{equation}
D_m[N^c] = \int d^3x \delta N^c \barpi \partial_c \delta \varphi,
\end{equation}
\begin{eqnarray}
H^Q_m[N] &=& \int d^3x \barN  \times \left[\mathcal{H}^{(0)}_{\pi} +\mathcal{H}^{(0)}_{\varphi}  
+\mathcal{H}^{(2)}_{\pi} + \mathcal{H}^{(2)}_{\nabla} + \mathcal{H}^{(2)}_{\varphi}  \right]\nonumber\\ 
&&+\int d^3x \delta N  \times\left[ \mathcal{H}^{(1)}_{\pi} + \mathcal{H}^{(1)}_{\varphi}  \right],
\end{eqnarray}
with
\begin{eqnarray} \lb{constraintholo}
\mathcal{H}^{(0)}_G&=& -6 \sqrt{\bar{p}} (\mathbb{K}[1])^2, \\
\mathcal{H}^{(1)}_G  &=& -4\sqrt{\bar{p}} \left(  \mathbb{K}[s_1] +\alpha_1\right)
\delta K^d_d  \\
&& -\frac{1}{\sqrt{\bar{p}}} 
\left(  \mathbb{K}[1]^2+\alpha_2  \right) \delta E^d_d +\frac{2}{\sqrt{\bar{p}}} (1+\alpha_3) \partial_c \partial^d \delta E^c_d,  \nonumber  \\
\mathcal{H}^{(2)}_G & =& \sqrt{\bar{p}}(1+\alpha_4) \delta K^d_c \delta K^c_d 
-\sqrt{\bar{p}} (1+\alpha_5)(\delta K^d_d)^2 \nonumber \\
&&-\frac{2}{\sqrt{\bar{p}}} \left(  \mathbb{K}[s_2]+\alpha_6 \right) \delta E^c_d \delta K^d_c   
\nonumber \\
&&-\frac{1}{2\bar{p}^{3/2}} \left( \mathbb{K}[1]^2+\alpha_7\right)\delta E^c_d \delta E^d_c \nonumber \\
&&+\frac{1}{4\bar{p}^{3/2}}\left( \mathbb{K}[1]^2+\alpha_8\right) (\delta E^d_d )^2 \nonumber \\
&&  - \frac{1}{2\bar{p}^{3/2}}(1+\alpha_9) \delta^{jk} (\partial_c \delta E^c_j)(\partial_d \delta E^d_k),  \lb{constraintholo3}
\end{eqnarray}
where we have set
\begin{equation}\lb{KK}
\forall n \neq 0, \hskip 0.3truecm \mathbb{K}[n] = \frac{\sin [n \mu \gamma \bark]}{n \mu \gamma},
\end{equation}
and $\mathbb{K}[n] = \bark$ for $n=0$. The $\K{s_i}$ are the holonomy-corrected versions of the single $\bark$ in the constraint, with $s_i$ an unknown parameter: this parametrization comes from the fact that when one quantizes using the holonomies, the field strength of the connection, initially appearing as $\bark^2$, becomes exactly $\K{1}^2$. Therefore, all the terms in the constraints that are in $\bark^2$ will be replaced by $\K{1}^2$, and because we do not know exactly the general correspondence $\bark \rightarrow \K{s_i}$, except when for squared terms, we will let $s_i$ be unknown integers. \\

In our approach, the integers $s_i$ are set by hand without any requirements except the one that at the end, for a specific value (or not) of $s_i$, the algebra should be closed \cite{Bojowald:2007cd}. However, it should be mentioned that there exists another approach trying rather to extract the effects of the quantum modifications directly from the quantum formalism by using an hybrid formalism: the Hybrid quantization \cite{superselection_sector}. In this other approach, generally, the quantization procedure is adapted by setting the value of the integer $s_i$, determining the step of displacement of the holonomies, as a result of the necessity to respect the superselection sectors of LQC.

Nevertheless, it has been noticed in  previous works, as explained in the former section, that one of the effects of the counterterms as used here and also added by hand, is to remove the terms depending on the parameters $s_i$ such that at the end, no dependence on these integers remains. This effect will be shown below.

We assume that 
\begin{equation}\label{muscheme}
\mu\propto \barp^\omega,
\end{equation}
for some constant $-\frac{1}{2}\leq\omega\leq0$. The case
$\omega=-\frac{1}{2}$ is  the so-called $\bar{\mu}$-scheme, whereas $\omega=0$ corresponds to the $\mu_0$-scheme. 

$\mathcal{H}^{(0)}_{\pi}$, $\mathcal{H}^{(0)}_{\varphi}$, $\mathcal{H}^{(1)}_{\pi}$, $\mathcal{H}^{(1)}_{\varphi}$, $\mathcal{H}^{(2)}_{\pi}$, $\mathcal{H}^{(2)}_{\nabla}$ and $\mathcal{H}^{(2)}_{\varphi}$ are not affected by the holonomy corrections \footnote{ which is not true when one considers the polymer quantization approach \cite{polymer_quant}.} and are therefore given by Eqs. (\ref{constmatter})-(\ref{constmatterEnd}).

\subsection{counterterms}

The associated counterterms are:

\begin{eqnarray}
\alpha_1 &=& \K{2}-\K{s_1},\\
\alpha_2 &=& 2 (\K{1}^2-\bark \K{2}) \lb{alpha2holo},\\
\frac{\partial \alpha_3}{\partial\bark} &=& 0, \label{alpha3} \\
\alpha_4 &=& -1+\text{cos}(2 \bark \gamma \barmu), \\
\alpha_5 &=& -1+\text{cos}(2 \bark \gamma \barmu), \\
\alpha_6 &=&-\bark \text{cos}(2 \bark \gamma \barmu)+2\K{2}-\K{s_2},\\
\alpha_7 &=&-2 \bark^2 \text{cos}(2 \bark \gamma \barmu)-4\K{1}^2+6\bark \K{2}, \\
\alpha_8 &=&-2 \bark^2 \text{cos}(2 \bark \gamma \barmu)-4\K{1}^2+6\bark \K{2}, \\
\alpha_9 &=&  \alpha_3 + 2 \barp \frac{d\alpha_3}{d\barp} .
\end{eqnarray}
Because of Eq. (\ref{alpha3}), $\alpha_3$ is an unknown function of $\barp$. 
\begin{eqnarray}
&& \beta_i= 0 \quad \text{for} \quad i\neq9,\\
&&\beta_9 = -1+ \text{cos}(2 \bark \gamma \barmu) (1+\alpha_3).
\end{eqnarray}

We find that
\begin{equation}
\Omega=\cos(2 \bark \gamma \barmu)(1+\alpha_3)\label{Omegaholo}.
\end{equation}
If one chooses $\alpha_3 = 0$, then necessarily $\alpha_9 = 0$, and the previous results \cite{scalars} are recovered, \textit{i.e} a possible change from an hyperbolic to an elliptic regime and/or an asymptotic silence scenario. Moreover, and this has not been underlined before, one can also keeps $\alpha_3$ and $\alpha_9$. As an heuristic example, let us consider the specific case $\alpha_9 = 0$. This leads to 
\begin{equation}
\alpha_3 = \frac{\text{constant}}{\sqrt{\barp}},
\end{equation}
and different cases must be considered:
\begin{itemize}
\item $constant = 0$, this has been extensively studied in  \cite{scalars}.
\item $constant>0 $, this is basically similar but with a modified dynamics. The Euclidean phase is dynamically shifted. The $\Omega$ term can be negative and even smaller than -1.
\item $constant<0$. When adjusting the value of the constant in a tuned way, this can lead to an evolution without the Euclidean phase. However, the asymptotic silence surface remains.
\end{itemize}
Numerical investigations have shown that the corresponding power spectrum of tensor perturbations does not depend heavily on the value of the $constant$. The shapes of the different power spectra are quite equivalent to the case $\alpha_3 = \alpha_9 = 0$. Too much freedom is however still remaining and more constraints are required to fully fix the dynamics.

\section{Inverse-volume and holonomy corrections considered simultaneously}
 
In the previous sections, the anomalies appearing when one considers either the holonomy or the inverse-volume correction alone were described. In this section, both terms are considered simultaneously and the resulting algebra is studied.

\subsection{constraints}

The constraints considered in this section are Eqs. (\ref{DG}), (\ref{conttot}), (\ref{Dm}) and (\ref{Hm})
but with $\mathcal{H}^{(0)}$, $\mathcal{H}^{(1)}$ and $\mathcal{H}^{(2)}$ given by Eqs. (\ref{constraintholo}) - (\ref{constraintholo3}). The matter Hamiltonian does not depend on the Ashtekar connection and is therefore not subject to holonomy corrections, except for $\beta_9$.

For the same reasons as explained before, we are allowed to choose $\gamma_3=\gamma_4=\gamma_5=\gamma_6=\gamma_7=\nu_1=\nu_3=\nu_4=0$ and $\sigma_0=\nu_0$ without any loss of generality.\\

\subsection{$\mu$-scheme}
In Section \ref{HOL} we have recalled the results of \cite{scalars} where $\mu$ was assumed to depend on $\bark$ according to $\mu\propto\barp^\omega$. In that case it was shown that we must have $\omega=-\frac{1}{2}$.

This time, we will try to be one step more general. We assume that $\mu$ is an unknown function of $\bark$ and define
\begin{equation}
\omega:=\frac{\barp}{\mu}\frac{d\mu}{d\barp}.
\label{smallomega}
\end{equation}
This definition is a generalization, and therefore is in agreement with the one previously used, $\mu\propto\barp^\omega$.

Since $\mu$ is an unknown function of $\barp$, so is $\omega$. However, we still expect $-\frac{1}{2}\leq\omega\leq0$, for the same reasons as before.

\subsection{Solution after closing the algebra}

The Poisson brackets will have the same shape as in the inverse-volume case, but with some modifications in the expressions of the anomalies, because of the holonomy corrections. However, no new anomaly arises, and the calculations are very similar.

Again, we find 
\begin{eqnarray}
&&\gamma_1=\gamma_2=\gamma_8=0,  \\
&&\nu_2=\nu_3=\nu_5=\nu_6=\nu_7=\nu_8=0,
\end{eqnarray}
and
\begin{equation}
\frac{\partial\nu_0}{\partial\bark}=0.
\end{equation}

As before, in order to simplify the equations, we  define:
\begin{eqnarray}
\Gamma_{\bark[1]} &:=& \frac{\K{1}}{\gamma_0}\frac{\partial \gamma_0}{\partial \bark}, \\
\Gamma_{\bark[2]} &:=& \frac{\K{2}}{\gamma_0}\frac{\partial \gamma_0}{\partial \bark}, \\
\Gamma_{\bark[1],2} &:=& \frac{\K{1}^2}{\gamma_0} \frac{\partial^2 \gamma_0}{\partial \bark^2}.
\end{eqnarray}
The counterterms for the gravity sector are given by:
\begin{eqnarray}
\alpha_1 &=&-\K{s_1} +\left(2\K{2}+\K{1}\Gamma_{\bark[1]}\right)\Sigma_{\barp},\\
\alpha_2 &=& 2 \K{1}^2-2 \bark \left(2\K{2}+\K{1}\Gamma_{\bark[1]}\right)\Sigma_{\barp},\\
\alpha_3 &=& -1 + \frac{f_1[\barp]}{\gamma_0},\\
\alpha_4 &=& -1+ \left(2 \text{cos} (2 \gamma \mu \bark) + 4 \Gamma_{\bark[2]} + \Gamma_{\bark[1],2} \right)\Sigma_{\barp},\\
\alpha_5 &=& -1+\left(2 \text{cos} (2 \gamma \mu \bark) + 4 \Gamma_{\bark[2]} + \Gamma_{\bark[1],2} \right)\Sigma_{\barp},\\
\alpha_6 &=& -\K{s_2}+ 2 \K{2} +\K{1}\Gamma_{\bark[1]} \nonumber \\
&& -\bark\left(2 \text{cos} (2 \gamma \mu \bark) + 4 \Gamma_{\bark[2]} + \Gamma_{\bark[1],2} \right)\Sigma_{\barp},\\
\alpha_7 &=& 6\bark \K{2} -\K{1}^2 (4+2\Lambda_\gamma)+2\bark \K{1}\Gamma_{\bark[1]} \nonumber \\
&& +2(1+2\omega)\left(\K{1}^2-\bark \K{2}\right)\nonumber \\
&& -2 \bark^2 \left(2  \text{cos}(2 \gamma \mu \bark) + 4 \Gamma_{\bark[2]} + \Gamma_{\bark[1],2} \right)\Sigma_{\barp},\\
\alpha_8 &=& 6\bark \K{2} -\K{1}^2 (4+2\Lambda_\gamma)+2\bark \K{1}\Gamma_{\bark[1]} \nonumber \\
&& +2(1+2\omega)\left(\K{1}^2-\bark \K{2}\right)\nonumber \\
&& -2 \bark^2 \left(2  \text{cos}(2 \gamma \mu \bark) + 4 \Gamma_{\bark[2]} + \Gamma_{\bark[1],2} \right)\Sigma_{\barp},\quad\\
\alpha_9 &=& -1 + \frac{1}{2 \gamma_0 \Sigma_{\barp}} \left(f_1[\barp] + 2 \barp \frac{df_1[\barp]}{d\barp} \right).
\end{eqnarray}

For the matter sector, without assumptions, the results are the same as for the inverse-volume case, except for $\beta_9$  which reads here as  \begin{equation}
\beta_9 = -1 + \frac{3 \gamma_0}{\nu_0^2} \frac{(2 \text{cos} (2 \gamma \mu \bark) + 4 \Gamma_{\bark[2]} + \Gamma_{\bark[1],2} )}{(3 - \Lambda_{\nu})}\Sigma_{\barp}f_1[\barp].
\end{equation}
 
All the counterterms have the correct ({\it i.e.} vanishing) classical limit.

The structure function of the algebra is now given by 
\begin{eqnarray}
\Omega &=& \gamma_0 \left(2 \text{cos} (2 \gamma \mu \bark) + 4 \Gamma_{\bark[2]} + \Gamma_{\bark[1],2} \right)\Sigma_{\barp} f_1[\barp]\nonumber\\
&=&\left(\frac{\partial^2}{\partial\bark^2} (\gamma_0\K{1}^2)\right)\Sigma_{\barp} f_1[\barp].\lb{OmegaIVHolo}
\end{eqnarray}

When both corrections are taken in to account, Eq. (\ref{prerest}) can be expressed as:
\begin{equation}\lb{Rest}
-\frac{\partial}{\partial\bark} \frac{\frac{\partial}{\partial\barp}\left(\frac{\gamma_0}{\barp}\K{1}^2\right)}{\frac{\partial}{\partial\bark}\left(\frac{\gamma_0}{\barp}\K{1}^2\right)}
=\frac{\partial}{\partial\barp} \ln\left(\sqrt{\barp}\ \Sigma_{\barp}\right),
\end{equation}
which can be shown to be equivalent to Eq. (\ref{rest}) when $\K{i}\to\bark$ ($\mu \to 0$).

If $\gamma_0=\nu_0=1$ then Eq. (\ref{Rest}) reduces to $\omega=-\frac{1}{2}$, and we recover the $\bar{\mu}$-scheme which was found as the solution in the holonomy case.

\subsection{Solutions after further assumptions}

\subsubsection{Case where $\gamma_0=\gamma_0[\barp]$}
If we assume that $\gamma_0$ is only a function of $\barp$ and not of $\bark$, Eq. (\ref{Rest}) becomes:
\begin{equation}
\frac{\Lambda_\gamma-1-2\omega}{\text{cos}(\gamma\bar{\mu}\bark)}=(3\Lambda_\gamma-1-2\omega)+4\barp\frac{\partial}{\partial\barp} \ln\left(\Sigma_{\barp}\right).
\end{equation}
The right hand side of the above equation is clearly $\bark$-independent, therefore the left hand side must be so too. Nevertheless, this is only the case if 
\begin{equation}\label{la}
\Lambda_\gamma=1+2\omega,
\end{equation}
and consequently 
\begin{equation}\label{si}
\barp\frac{\partial}{\partial\barp} \ln\left(\Sigma_{\barp}\right) = -\frac{1}{2}\Lambda_\gamma.
\end{equation}

Eqs. (\ref{la}) and (\ref{si}) have the solutions
\begin{eqnarray}
\mu&\propto&\sqrt{\frac{\gamma_0}{\barp}},\\
\Sigma_{\barp}&=&\frac{1}{2\sqrt{\gamma_0}},
\end{eqnarray}
when the correct classical limit has been taken in to account.

From this, we see that if the inverse-volume corrections are unaffected by the holonomy corrections, then the holonomy corrections will be affected by the inverse-volume corrections. Therefore we should conclude that either the inverse-volume corrections are affected by the holonomy corrections, or the holomomy corrections are affected by the inverse-volume corrections.

\subsubsection{Case where $\omega=-\frac{1}{2}$ and $\gamma_0=\gamma_0\left[\frac{\bark}{\sqrt{\barp}}\right]$}
In the case where $\omega=-\frac{1}{2}$ and $\gamma_0=\gamma_0\left[\frac{\bark}{\sqrt{\barp}}\right]$,  $\frac{\gamma_0}{\barp}\K{1}^2$ is now a function of $\frac{\bark}{\sqrt{\barp}}$. Eq. (\ref{Rest}) reduces thus to $\Sigma_{\barp}=\frac{1}{2}$ which is exactly the result obtained previously, without considering the holonomy corrections, leading $\nu_0$ to evolve as in Eq. (\ref{solnu}).\\
The structure function is now given by 
\begin{equation}
\Omega=\left(\gamma_0\text{cos}(2\gamma\bar{\mu}\bark)+2\frac{\K{2}}{\sqrt{\barp}}\gamma_0'+\frac{1}{2}\frac{\K{1}^2}{\barp}\gamma_0''\right)f_1[\barp].
\end{equation}
In this case, $\Omega=\Omega\left[\frac{\bark}{\sqrt{\barp}}\right]$ if and only if $f_1[\barp]\equiv1$.

\subsubsection{Case where $\Omega=\Omega[\barp]$}
In the same way as before, if $\Omega$ does not depend on $\bark$, then Eq. (\ref{OmegaIVHolo})
can be solved for $\gamma_0$ giving in the considered case:
\begin{equation}
\gamma_0=\frac{1}{\K{1}^2}\left(\frac{\bark^2\Omega}{2\Sigma_{\barp}f_1[\barp]}+\bark f_2[\barp]+f_3[\barp]\right).
\end{equation}
It is interesting to notice that the expression of the correction has a really similar shape as the one given by Eq. (\ref{solgamma}) for the inverse-volume correction. Moreover, as said previously, in order to understand the action of the correction and its characteristics on the related phenomenology, it is necessary to consider it with the constraints densities. An interesting consequences arises : in the gravitational constraint density, the term in the numerator is now given by $\K{1}^2$ which cancels exactly the same term in the correction, and only the terms in $\bark$ remains. In other words, after including this correction in the constraint, the final constraint will have exactly the same expression as the one derived previously for the inverse-volume correction only. Therefore, this case is not interesting, the holonomy correction being cancelled by the inverse-volume correction, even if the idea of the bounce remains. One can consequently assume that the interesting cases are obtained in a similar way as when the holonomy correction was considered, that is to say when the structure function $\Omega$  depends not only on $\barp$, but also on $\bark$. This question is left opened for further investigations.

\subsection{Some remarks on the case where both corrections are taken into account}

%Let us here briefly summarize some of the assumptions made so far:
%\begin{enumerate}
%\item the counterterms do not depend on the matter content,
%\item the holonomy correction is not affected by the inverse-volume correction: the $\barmu$-scheme is used $\beta = - \frac{1}{2}$,
%\item the inverse-volume correction has to depend on the connection,
%\item the assumption $\beta_9 = \beta_{10}$ emphasizes the idea of a close similarity between $\beta_1$ and $\beta_3$, and leads to some rather interesting even if pessimistic fact that the constraints are similar to the ones of general relativity.  
%\end{enumerate}

We have shown that the full algebra can be closed and that either the inverse-volume correction has to depend on the holonomy one, or vice-versa.\\

It should also be noticed that Eq. (4.2) and (4.3) of \cite{thiemann} read respectively
\begin{equation}
E^j_k = \frac{2}{\kappa^2} \int \epsilon_{krs} \epsilon^{abc} \{ A^r_b,V(R)\}\{A^s_c,V(R)\} n^S_a,
\end{equation}
and 
\begin{equation} \lb{theoeq}
E^j_k \sim Tr[ h \{ h^{-1}, V\} \tau_k h  \{ h^{-1}, V\} ].
\end{equation}
Considering the quantum version of the previous equation, one can hope to gain a better understanding of the most general inverse-volume correction and its interplay with the holonomy correction. This is however beyond the scope of this article.

\section{Gauge invariant variables and equations of motion}

In order to understand better the consequences of the counterterms presented in this work, it is interesting to study the modifications they induce on the equations of motion for the cosmological perturbations. This section is therefore dedicated to the derivation of the generic equations of motion, for scalar and tensor modes, by first expressing them with the general expressions for the counterterms, and then by replacing in the final equations these expressions by the ones found in the previous sections. In the following, only some details will be given as one can simply follow what has been done in \cite{Bojowald:2008jv} and \cite{notrepapierHamJac} for instance. However a reader having a specific model in mind will be able to compute to the end the corresponding equations of motion.
\\

As in \cite{consistency}, one can write the general perturbation of the densitized triad as
%\com{(This part is from the article with Francesca, "consistency...")}
\begin{eqnarray}
\delta E^a_i &=& \barp \;\Big[ - 2 \psi \delta^a_i + (\delta^a_i \partial^d \partial_d  - \partial^a \partial_i )E 
\nonumber\\ && 
- c_1 \partial^a F_i - c_2 \partial_i F^a - {\scriptstyle\frac{1}{2}} h^a_i \; \Big],
\end{eqnarray}
where the first two terms $\psi$  and  $E$ correspond to the scalar modes, $F_i$ and $F^a$ to the vector modes, and $h_i^a$ to the tensor modes. One will  also have to deal with the perturbations of the matter content, which are related to the perturbations of the metric by the Einstein equations.\\
 Moreover, expressing the perturbations of the metric as in the previous equation (choosing a specific gauge) will also constrain the lapse function $\delta N$ and the shift vector $\delta N^a$. The form of the metric in the case of vector and tensor modes implies that the variation of the lapse is zero: $\delta N=0$\,. For vector modes, the variation of the shift corresponds to one of the two degrees of freedom: $\delta N^a = S^a$, while for the tensors modes, $\delta N^a=0$. One can then notice that for the tensor modes, due to the vanishing of the perturbations of the lapse function and of the shift vector, the first order in the constraints densities will never be considered. \\
Vector modes are transverse, and tensor modes are transverse and traceless. These conditions constrain $\delta E^a_i$ and  $\delta K^i_a$. In particular, the vanishing trace implies
\begin{equation}
\delta^i_a \delta E^a_i =\delta ^a_i \delta K^i_a= 0,
\end{equation}
which, when one considers the case of the tensor or vector perturbations, leads to a simplification of the expressions of the constraints, the terms proportional to the traces above vanishing. Moreover, for the tensor modes, the transverseness, {\it i.e.} vanishing divergence, implies also 
\begin{equation}
\partial^ i\delta E^a_i =
\partial_a \delta E^a_i = 0~.
\end{equation}

The more general case is therefore given by the one for the scalar perturbations, no term disappearing due to the properties of the perturbations, with consequently a contribution of all orders of the constraints densities. In this case, we have
\begin{equation} \lb{pertscaldeltaN} \!
 \delta N = \barN \phi \hskip 5mm \mbox{and}  \hskip 5mm \delta N^a = \partial^a B, \!\!
\end{equation}
where $\barN$ is the unperturbed part of the lapse $N=\barN+\delta N$, and $\phi$ and $B$ are the other degrees of freedom for the scalar perturbations in the metric : considering only the metric, only 2 of these degrees of freedom are physically relevant which, with the Einstein equations, reduce to one when considering also the perturbations of the matter content, the remaining true physical degree of freedom is the Mukhanov-Sasaki variable, $v$, as explained later.

\subsection{Background equation of motion}
Because we neglect the backreaction of the perturbations on the background, its homogeneous part on which the perturbations live, is not affected by the next orders from which the previous counterterms have been derived. However it will be affected by the zeroth order of the inverse-volume and holonomy corrections. In this section, we give the equations of motion for the background, namely the Friedmann and Klein-Gordon equations, when both corrections are considered simultaneously. One can recover the decoupled equations by taking the corresponding limits. 

The Friedmann equation, partially derived first in \cite{Calcagni:2008ig}, is given by 
\begin{equation}
H^2=\frac{\kappa}{3}\gamma_0\rho\left(1-\frac{\rho}{\rho_c \gamma_0}+\Gamma_{\bark[2]}+\frac{1}{4}\Gamma_{\bark[1]}^2\right),
\end{equation}
where
\begin{equation}
\rho:=\frac{1}{2\nu_0}\left(\frac{d\bar{\varphi}}{dt}\right)^2 +V(\bar{\varphi}),\quad \rho_c:=\frac{\kappa}{3\gamma^2\mu^2\barp},
\end{equation}
and the Klein-Gordon equation for the matter field by
\begin{equation}
\frac{d^2\bar{\varphi}}{dt^2}+(3-2\Lambda_\nu)H\frac{d\bar{\varphi}}{dt}+\nu_0 V'(\bar{\varphi})=0.
\end{equation}
Here, $H$ is the Hubble factor in cosmic time $t$.

\subsection{Tensor variables and equations of motion}
In this section, we derive the corresponding equations of motion for the tensor variables $h_{\mu \nu}$ which are gauge-invariant. The two degrees of freedom will be called in the following $h = h_\times = h_+$. We consider here the case where both corrections are taken into account simultaneously. To make the expression clearer, we will firstly not fix the counterterms. 

After some algebra following \cite{Bojowald:2007cd}, one obtains  the expression of $\delta K^i_a$,
\begin{equation}
\gamma_0 (1+\alpha_4) \delta K^i_a = \frac{1}{2} \left( \dot{h} + h \gamma_0 (2 \K{2} - \K{s_2}-\alpha_6 ) \right),
\end{equation}
 whose equation of motion leads to the equation of motion for the tensor perturbations:
\begin{eqnarray} \lb{eqtens}
&0&= \ddot{h} + \dot{h} \left(\frac{\partial}{\partial\bark}\left(\gamma_0 \K{1}^2\right) - \frac{1}{\gamma_0 (1+\alpha_4)} \frac{d }{d \eta}(\gamma_0 (1+\alpha_4)) \right) \nonumber \\
&&-\gamma_0^2 ( 1 + \alpha_4)(1+\alpha_9) \partial_i \partial^i h \nonumber \\
&& + h \left[ \gamma_0 (\K{s_2}+ \alpha_6)\left(\frac{\partial}{\partial\bark}\left(\gamma_0 \K{1}^2\right)- \gamma_0 (\K{s_2}+ \alpha_6) \right)  \right. \nonumber\\
&&\left. + \frac{d}{d \eta} \left(\frac{\partial}{\partial\bark}\left(\gamma_0 \K{1}^2\right)- \gamma_0 (\K{s_2}+ \alpha_6) \right)  \right. \nonumber \\
&& \left. - \left(\frac{\partial}{\partial\bark}\left(\gamma_0 \K{1}^2\right)- \gamma_0 (\K{s_2}+ \alpha_6) \right) \frac{\frac{d}{d\eta} (\gamma_0 (1+\alpha_4))}{\gamma_0 (1+\alpha_4)}\right. \nonumber \\
&& \left. - \frac{1}{2} \gamma_0^2 (1+\alpha_4) \left(\K{1}^2 + \alpha_7\right) \right. \nonumber  \\
&& \left. + \frac{\kappa}{2} \gamma_0 (1+\alpha_4) \left( (1+\beta_8) \frac{\barnu \barpi^2}{2 \barp^2} - \barp V (1+\beta_{13})\right) \right]. 
\end{eqnarray}
In the above equation we have used $\gamma_i=\nu_i=0$ for $i\neq0$, and also $\frac{d\nu_0}{d\bark}=0$, but left the counterterms $\alpha_j$ and $\beta_j$ general. \\
Surprisingly, when the expression of the counterterms derived previously are inserted into the former equation, one can derive a really simpler expression  
\begin{eqnarray}
0 &=& h'' + h' \left( 2\Hc\left(1+\frac{2\barp}{f_1[\barp]}\frac{df_1[\barp]}{d\barp}\right) -\frac{{\Omega}'}{\Omega}\right) \nonumber \\
&&-\frac{\Omega}{2\Sigma_{\barp}}\left(1+\frac{2\barp}{f_1[\barp]}\frac{df_1[\barp]}{d\barp}\right)\nabla^2 h, \lb{tensor}
\end{eqnarray}
where the prime refers here to the conformal time $\eta$, $\Hc$ is the conformal Hubble factor  $\Hc=\frac{{a}'}{a}=\frac{{\barp}'}{2\barp}$, and $\Omega$ is given in the general case by Eq. (\ref{OmegaIVHolo}). When one  considers only the inverse-volume correction, $\Omega$ is given by Eq. (\ref{OmegaIV}) whereas for the holonomy correction it is given by Eq. (\ref{Omegaholo}) with also $f_1[\barp]\to(1+\alpha_3)$ (the dependence on $\alpha_3$ was missed in previous works, since $\alpha_3$  was there assumed to be zero). The equation of motion can also be simplified when one considers $\Omega = 1$, as for instance in Eq. (\ref{omegap}).\\
Moreover, it is interesting to notice that the tensor perturbations propagate with the speed $c_t$ given by
\begin{equation}
c_t^2=\frac{\Omega}{2\Sigma_{\barp}}\left(1+\frac{2\barp}{f_1[\barp]}\frac{df_1[\barp]}{d\barp}\right).
\end{equation}
In conclusion, we can see that there is no term proportional to $h$, which is usually expected at the classical limit, but also when the holonomy correction only is considered. In order to investigate the consequences of the inverse-volume correction, for instance, a possible study could be performed with $\Omega = 1$ and Eq. (\ref{omegap}), such that the expression of the inverse-volume correction could be expressed as Eq. (\ref{evolIV}). This work, with the help of \cite{jakub_cosmo}, could be compared with what has been derived in \cite{Bojowald:2011hd}. This is nevertheless left for future investigation.

\subsection{Scalar variables and equations of motion}

In this section, the scalar gauge-invariant variables and their equations of motion are derived for a general expression of the counterterms. The procedure, conventions and notations used here are described in details in \cite{notrepapierHamJac}, based on \cite{Langlois}, but some steps will be recalled briefly in the following (the reader interested  in the way the equations are derived is invited to read the previously cited articles). 

The calculations are quite tedious in some cases. To simplify, we first assume that the gauge-invariant variable has the "minimal required shape" $Q$ related to the Mukhanov-Sasaki variable $v$: for instance, classically, $ Q = \frac{v}{\sqrt{\barp}}= \delta \varphi - \frac{\dot{\bar{\varphi}}}{\sqrt{\barp} \mathcal{H}} \psi $ where $ \delta \varphi $ and $\psi$ are respectively the first order perturbations for the matter (a single Klein-Gordon field $\bar{\varphi}$) and for the metric, with $\mathcal{H}$ the conformal Hubble parameter described previously. As  in \cite{notrepapierHamJac}, for simplicity, the calculation of the Hamiltonian giving the equations of motion for the gauge-invariant variables will be  first done using the variable $Q$, and then the usual Mukhanov-Sasaki action will be rederived for the variable $v$. 
As shown in the following, in order to switch from the Hamiltonian to the Lagrangian, the second Mukhanov-Sasaki variable $z$ will have to fulfill classically the relation
\begin{equation}\lb{relationconsistency}
\frac{\ddot{z}}{z} + \Gamma - k^2 -\left(
\frac{1}{\sqrt{\barp}}\frac{d(\sqrt{\barp}) }{d\eta}\right)^2 -
\frac{d}{d\eta} \left( \frac{1}{\sqrt{\barp}}\frac{d(\sqrt{\barp})
}{d\eta}\right) = 0,
\end{equation}
where $\Gamma$ corresponds to the effective mass of $Q$ in the Hamiltonian and $k^2$ is the eigenvalue of the Laplacian operator after a Fourier transform. It should be noticed that contrarily to the background variables where the Hamiltonian
is in fact a constraint, for the gauge invariant variable it
corresponds to a true Hamiltonian.

In this procedure, in order to perform a change of variables from the different perturbations to the gauge invariant ones, as $Q$ or $v$, one defines a symmetric matrix $A_{ij}$, which by definition allows $Q$ and $v$ to commute with the constraints as shown in \cite{notrepapierHamJac}, and whose components will affect the effective mass $\Gamma$ of the gauge invariant variables. The important components, $A_{00}$ and $A_{01}$, can be calculated and are given by:
\begin{equation}
A_{00} = -\frac{\barp}{2 \nu_0 (1+\beta_5)} \left(  \frac{\dot{\bar{\varphi}}^2}{\nu_0} \frac{(1+\alpha_4)(1+\beta_1)}{(\K{s_1}+\alpha_1)} \right),
\end{equation}
and
\begin{eqnarray}
A_{01} &=& - \frac{1}{2 \barp \gamma_0 (\K{s_1}+\alpha_1)} \left[ A_{00} (1+\beta_1) \dot{\bar{\varphi}} + \barp^2 (1+\beta_3)\partial_{\varphi } V\right], \nonumber 
\end{eqnarray} 

such that $\Gamma$ is 

\begin{eqnarray}
\Gamma &=& f(A_{ij}, \barp, \bark,...) (\alpha_4 -\alpha_5) + \frac{\dot{A}_{00}}{\barp} + \frac{3}{2} \gamma_0  (1+\alpha_4)\left(\frac{\dot{\bar{\varphi}}}{\nu_0} \right)^2 \nonumber \\ 
&&- 2  \gamma_0 (1+\alpha_4)\frac{\dot{\bar{\varphi}}}{\nu_0} A_{01} +\nu_0 (1+\beta_5) \left(\frac{A_{00}}{\barp} \right)^2 \nonumber \\
&& + \nu_0 (1+\beta_9) k^2+\barp (1+\beta_{10}) \partial_{\varphi \varphi} V.
\end{eqnarray}
Due to the modification of the constraints with the counterterms, we need to redefine $v$ such that at the end
\begin{equation}\lb{defvmodif}
v = \sqrt{\frac{\barp}{\chi}} Q=\sqrt{\frac{\barp}{\chi}}  \left( \delta \varphi + \frac{\dot{\bar{\varphi}}}{\gamma_0} \frac{(1+\beta_1)}{(\K{s_1}+\alpha_1)} \Psi\right),
\end{equation}
where 
\begin{eqnarray}
\chi &=& \nu_0 (1+\beta_5)\nonumber\\
&=&\frac{\nu_0^2 (3-\Lambda_\nu)^2}{9}.
\end{eqnarray}
The Hamiltonian for $Q$ is then given by  
\begin{equation}
H^S_{GI} = \int \frac{d^3k}{2} \left[\left(\sqrt{\frac{\chi}{\barp}} P \right)^2+(\chi \Gamma) \left(\sqrt{\frac{\barp}{\chi}} Q \right)^2 \right].
\end{equation}
With the definition of $v$ given in Eq. (\ff{defvmodif}), it is possible to obtain the usual Mukhanov-Sasaki action
\begin{equation}
\mathcal{S} = \int d\eta \int d^3k \frac{1}{2} \left[ \dot{v}^2 + \left( -c_s^2 \, k^2 + \frac{\ddot{z}}{z}\right) v^2 \right],
\end{equation}
such that $z$ has to fulfill a similar equation than Eq. (\ff{relationconsistency}), now modified due to the counterterms:
\begin{eqnarray} \lb{relationconsistencygen}
0 &=& \frac{\ddot{z}}{z} +{\chi} \Gamma - \nu_0 (1+\beta_9){\chi} k^2 -\left( \left(\sqrt{\frac{\barp}{\chi}} \right)^{-1} \frac{d}{d\eta} \left(\sqrt{\frac{\barp}{\chi}} \right) \right)^2  \nonumber \\
&&  - \frac{d}{d\eta} \left( \left(\sqrt{\frac{\barp}{\chi}} \right)^{-1} \frac{d}{d\eta}\left( \sqrt{\frac{\barp}{\chi}}\right)\right) = 0. 
\end{eqnarray}
Previous works have suggested that $z$ would have in fact the following expression:
\begin{eqnarray}\lb{defz}
z &=& \sqrt{\frac{\barp}{\chi}} \frac{\dot{\bar{\varphi}}}{\gamma_0} \frac{(1+\beta_1)}{(\K{s_1}+\alpha_1)}\nonumber\\
%&=&\frac{\sqrt{\barp}\dot{\bar{\varphi}}\left(3-\Lambda_\nu\right)\Sigma_{\bark}}{3\nu_0\gamma_0\left(2\K{2}+\K{1}\Gamma_{\bark[1]}\right)},\\
&=& \frac{ \sqrt{p} \dot{\bar{\varphi}} }{ \nu_0 \gamma_0 ( 2\K{2} + \K{1}\Gamma_{\bark[1]} ) \Sigma_{\barp} }
\end{eqnarray}
but at this stage, due to some unknown relations and to the difficulties of the calculations, this expression remains only an assumption, which works classically and in the case of the holonomy correction.\\
The speed of propagation for the scalars is in general given by 
\begin{eqnarray}
c_s^2 &=& \nu_0 (1+\beta_9)\chi\nonumber\\
&=&\frac{\nu_0\left(3-\Lambda_\nu\right)}{3}\Omega
\end{eqnarray}
and it is interesting to notice that this expression of $c_s$ is different from the one $c_t$ derived for the tensor perturbations. This could be explained by the fact that $c_t$ depends on $\alpha_9$ in Eq. (\ref{eqtens}), and it was shown in \cite{consistency} that the corresponding term in the Hamiltonian constraint is in fact different when one considers the tensor perturbations. Therefore, the expression obtained here for $\alpha_9$ was in fact specific to the scalar perturbations and could be different for the tensor perturbations. Nevertheless, due to the properties of the tensor perturbations, the algebra for these kind of perturbations corresponds to the homogeneous one, where $\{H,H\}=0$, and it is not possible to derive an expression for the counterterms, which remain unknown if one considers the approach only for these perturbations. In order to solve this problem, one could also assume that $c_s = c_t$ and see what are the consequences on the phenomenology induced by the previous modifications (in a similar way as the case where only the holonomy corrections were considered, where $c_s=c_t=\Omega$ if and only if $\alpha_3=0$, and whose spectrum was derived in \cite{spectrum_linda}). This is also left for future works. \\
When one considers the holonomy correction only, the variables are then given by:
\begin{equation}
\chi = 1 \quad,\quad 
z = \sqrt{\barp}  \frac{\dot{\bar{\varphi}}}{\K{2}}, \quad \text{and}\quad c_s^2=\Omega,
\end{equation}
as found in \cite{scalars}. However, due to the unknown expressions, when the inverse-volume corrections are considered, we will not derive here the expressions of the equation of motion with the specific counterterms given previously, the expression being too complicated and not useful. Nevertheless, the reader who would have a specific set of counterterms will be able  to derive  without any ambiguities the corresponding equation of motion for the scalar perturbations and check that what has been said previously will work also in this case.

\section{Conclusions}

In this work, we have tried to sum-up all that is currently known on the issue of the closure of the algebra in effective LQC and to address the question of a full resolution taking into account both corrections simultaneously. To this aim, we have generalized the previously derived results for the inverse-volume corrections. We have also found new solutions, that were missed in previous works, for the holonomy case. 

We have found that in all three cases (holonomy, inverse-volume, and holonomy + inverse-volume) it is possible, under the assumptions of this paper, to close the algebra of constraints. Further, we have calculated the explicit counterterms required for this closure. These counterterms are functions of the zeroth order corrections and of an unknown function of integration ($f_1[\barp]$ in the case of inverse-volume or inverse-volume + holonomy, and $\alpha_3$ in the case of holonomy only).

An interesting result is that the final form of the constraints does only depend on the zeroth order of the corrections and not on higher order terms. \\

We have found some equations constraining the form of the corrections. This includes (not exhaustively):
\begin{itemize} 
\item in both the inverse-volume and inverse-volume + holonomy cases:
\begin{equation}\frac{\partial\nu_0}{\partial\bark}=0.\end{equation}
\item in the case with only holonomy corrections:
\begin{equation}\mu\propto\frac{1}{\sqrt{\barp}}.\end{equation}
\item in the case of both corrections simultaneously:
\begin{equation}\frac{\partial\gamma_0}{\partial\bark}\neq0 \qquad \text{or} \qquad \mu\propto\sqrt{\frac{\gamma_0}{\barp}}.\end{equation}
\end{itemize}

The inverse-volume corrections are constrained but far from being fully determined by the closure of the algebra. This is normal, and the final input is expected to come from the full theory.\\

We have found that the final algebra is modified when compared with the classical one, by a factor $\Omega$ given by
\begin{equation}
\Omega=\left(\frac{\partial^2}{\partial\bark^2} (\gamma_0\K{1}^2)\right)\Sigma_{\barp}f_1[\barp],
\end{equation}
in the case of both corrections. The inverse-volume case is given by $\barmu \to 0$ or $\K{n} \to \bark$ and the only holonomy case is given by $\gamma\to1$, $\Sigma_{\barp}\to\frac{1}{2}$ and $f_1[\barp]\to(1+\alpha_3)$.

If the holonomy corrections are included (with or without inverse-volume corrections), then a modification of the algebra should be unavoidable, that is, there is no solution such that $\Omega=1$ which does not completely cancels the holonomy corrections. Surprisingly, a bounce might still be possible whenever the inverse-volume corrections are  considered.

In the case of only inverse-volume corrections, the solution $\Omega=1$ is not excluded, showing an interesting example of an unmodified algebra with nevertheless some modifications of the constraints, in tension with a naive interpretation of the Hojman-Kuchar-Teitelboim theorem. 

Moreover, the equations of motion for the tensor and scalar perturbations have been derived in the general case, but a full example when both corrections are considered simultaneously is still awaited. Some toy models could also be studied, leading to some open new perspectives for phenomenology. 

To conclude, the simplest case for the constraints has been considered so far, and the next steps could be either to include  backreaction or  higher derivatives in the expressions of the constraints.

\acknowledgments
The authors would like to thank Martin Bojowald, Norbert Bordendofer but also the referees for their improvement of this article, by their careful readings and comments. T.C was supported by the NSF grant PHY-1205388. L.L. is supported by the Labex ENIGMASS.

\end{document}